\documentclass[12pt, a4paper]{article}
\usepackage[margin=2.5cm]{geometry}
\usepackage[english]{babel}
\usepackage[utf8]{inputenc}
\usepackage[T1]{fontenc}
\usepackage{titling} % for custom title page on article class

\usepackage{graphicx}
\graphicspath{{./Images/}}
\usepackage[bf]{caption} % caption: FIG in bold

\usepackage{setspace}

\usepackage[square, numbers, sort&compress]{natbib} % for bibliography - Square brackets, citing references with numbers, citations sorted by appearance in the text and compressed (as in [4-7])

\usepackage{amsthm,amsmath,amssymb,bbm}% Math symbols
\usepackage{tensor,mathtools,slashed,braket}% more symbols, tensor and dirac notation stuff
\usepackage{float,wrapfig} % wrapping text around figures
\usepackage[compat=1.1.0]{tikz-feynman} % feynman diagrams
\tikzfeynmanset{warn luatex = false} % disable luatex warning; need luatex for automatic vertex placement

\usepackage{hyperref}% hyperref adds clickable links at references, often should be loaded last

\usepackage{lipsum}

\newcommand{\mycomment}[1]{} % multiline comments
\newcommand{\email}[1]{\thanks{Electronic Address: \href{mailto:#1}{#1}}}

\newcommand{\tr}{\text{Tr}}

\begin{document}

\title{$\mathcal{N}=2$ supersymmetry in the twistor description of higher-spin holography}

\preauthor{\begin{center}
		\large \lineskip 0.5em%
		\begin{tabular}[t]{c}}
		
\author{Julian Lang\email{julian.lang.research@gmail.com} ~and Yasha Neiman\email{yashula@icloud.com}}

\postauthor{\vspace{0.5em}\\
	\textit{Okinawa Institute of Science and Technology},\\
	\textit{1919-1 Tancha, Onna-son, Okinawa 904-0495, Japan}%
	\end{tabular}\par\end{center}}

\date{\today}

\begin{titlingpage}
	\maketitle

	\begin{abstract}
		
		We study the holographic duality between higher-spin (HS) gravity in 4d and free vector models in 3d, with special attention to the role of $\mathcal{N}=2$ supersymmetry (SUSY). For the type-A bosonic bulk theory, dual to spin-0 fields on the boundary, there exists a twistor-space description; this maps both single-trace boundary operators and linearized bulk fields to spacetime-independent twistor functions, whose HS-algebra products compute all boundary correlators. Here, we extend this description to the type-B bosonic theory (dual to spin-$1/2$ fields on the boundary), and to the supersymmetric theory containing both. A key role is played by boundary bilocals, which in type-A are dual to the Didenko-Vasiliev $1/2$-BPS ``black hole''. We extend this to an infinite family of linearized $1/2$-BPS ``black hole'' solutions. Remarkably, the full supersymmetric theory (along with the SUSY generators) fits in the same space of twistor functions as the type-A theory. Instead of two sets of bosonic bulk fields, the formalism sees \emph{one} set of linearized fields, but with both types of boundary data allowed. 
		
	\end{abstract}
\end{titlingpage}

\tableofcontents
\newpage

%----------------------------------------------------------------------------------------
%	MAIN TEXT
%----------------------------------------------------------------------------------------

\onehalfspacing

\section{Introduction}

Higher-spin (HS) gravity \cite{VASILIEV1996,Vasiliev1990,VASILIEV2000} is the putative interacting theory of an infinite tower of massless gauge fields with increasing spin. It plays an important role in mathematical physics, being both a larger cousin of supergravity and a smaller cousin of string theory. Here, we'll be interested in three versions of the theory, all in 4 spacetime dimensions:
\begin{itemize}
	\item Type-A -- a bosonic theory with a parity-even master field $C(x;Y)$. This carries one field (a Weyl-curvature-like field strength) of each integer spin $s=0,1,2,\dots$, encoded as Taylor coefficients w.r.t. the twistor coordinate $Y$.
	\item Type-B -- a different bosonic theory with a parity-\emph{odd} $C(x;Y)$, but otherwise the same field spectrum.
	\item The $\mathcal{N}=2$ supersymmetric theory \cite{Vasiliev1990} whose bosonic sector is the union of type-A and type-B.  
\end{itemize}
HS gravity can be difficult to work with, since its original formulation -- the Vasiliev equations -- is very indirect (yet still viable and developing: Vasiliev's prescription has been recently amended \cite{Gelfond2018,Didenko2018,Didenko2019,Gelfond2020}, resolving an initially pathological behavior of the cubic vertices \cite{Boulanger2016}, and narrowing the initial functional freedom in the interactions to a single phase parameter). On the other hand, the theory has an extremely simple holographic description via AdS/CFT \cite{Maldacena1997,Aharony2000}: with boundary conditions that preserve HS symmetry, the holographic dual is just a free vector model in 3d \cite{Klebanov2002,Sezgin2003}. For type-A, this boundary dual is a $U(N)$ vector model of free spin-0 fields $\phi^I$ (with $I=1,\dots,N$ an internal index); for type-B, it is a $U(N)$ vector model of free spin-$1/2$ fields $\psi^I$; finally, for the $\mathcal{N}=2$ bulk theory, the boundary dual is the supersymmetric theory of both $\phi^I$ and $\psi^I$. Moreover, for type-A restricted to even spins, the holographic duality can be extended from AdS to de Sitter space, giving a rare \& precious working model of $dS_4/CFT_3$ (in fact, this is our main motivation for working on HS gravity). In this paper, we extend some recent results and techniques from the type-A case to the type-B and $\mathcal{N}=2$ theories. Though we'll work in Euclidean AdS space, our results will suggest some interesting new perspectives on the (traditionally purely type-A) de Sitter setting. 

We will deal with two seemingly unrelated aspects of the theory: AdS/CFT correlators, and BPS black-hole-like solutions. These are linked by the recent observation (in type-A) \cite{David2020a,Lysov2022a} that the linearized version \cite{Didenko2008} of the Didenko-Vasiliev BPS ``black hole'' \cite{Didenko2009} is nothing but the bulk dual of a \emph{single-trace bilocal operator} on the boundary, which in turn plays a key role in the HS-algebra structure of boundary correlators \cite{Neiman2017,Neiman2022}. 

Our working formalism will be the one developed for type-A in \cite{Neiman2017,David2020}. This in turn is based on a key observation in \cite{Didenko2012,Didenko2013} (for type-A and type-B respectively): that the boundary correlators can be expressed using the \emph{linearized} version of Vasiliev's bulk language, by taking HS-algebra products of linearized master fields $C(x;Y)$ that describe boundary-to-bulk propagators. The added insight of \cite{Neiman2017,David2020} is that, by using an appropriate embedding-space formalism, one can replace the bulk fields by their \emph{Penrose transforms} $F(Y)$, i.e. pure twistor functions with manifestly no spacetime dependence. In this formalism, which we review in section \ref{sec:basics}, both the bulk Penrose transform and the boundary Feynman diagrams are realized within HS algebra. The present paper sets out to apply this formalism to the type-B and $\mathcal{N}=2$ theories. We note that all of our results could have been seen already in the formalism of \cite{Didenko2012,Didenko2013}. The spacetime-independent twistor language of \cite{Neiman2017,David2020} is simply more covariant, and thus more transparent. 

Our two main results are as follows. First, we find that the $\mathcal{N}=2$ boundary operators \& correlators (including both the type-A and type-B bosonic sectors), as well as the $\mathcal{N}=2$ supersymmetry algebra, all fit within \emph{the same space of twistor functions $F(Y)$} as those of the type-A theory: there is no need for extra fermionic coordinates as in \cite{Fradkin1987,Konstein1990,Sezgin2012}, or for external Klein operators $k,\bar k$ as in \cite{Vasiliev1990,VASILIEV2000}. Specifically, the role of $k\bar k$ will be played by an \emph{internal} Klein operator -- the twistor delta function $\delta(Y)$. This suggests an unusual interpretation: if both the type-A and type-B theories sit in the same space of twistor functions, then they describe not two separate sets of fields $C(x;Y)$, but rather different configurations for the \emph{same} set of fields. We discuss this new interpretation and its implications in section \ref{sec:newbulkpicture}.

We now turn to our second main result. As mentioned above, in the type-A theory, the linearized Didenko-Vasiliev ``black hole'' is holographically dual to a bilocal boundary operator. As noted in \cite{Didenko2009}, despite being purely type-A, this solution is $1/2$-BPS with respect to the $\mathcal{N}=2$ supersymmetry (the original claim \cite{Didenko2009} was $1/4$-BPS, later corrected in \cite{Bourdier2015}). Here, we'll consider a more general class of boundary bilocals, both type-A and type-B, which correspond in the bulk to the linearized versions of certain generalizations of the Didenko-Vasiliev ``black-hole'' constructed in \cite{Iazeolla2011,Iazeolla2017}. Taking appropriate superpositions of these, we then find an infinite family of new $1/2$-BPS solutions (these solutions were not found in \cite{Bourdier2015}, which looked at more restricted generalizations of Didenko-Vasiliev). We note that our results are in discrepancy with the original Didenko-Vasiliev paper \cite{Didenko2009}: there, the type-A and type-B sectors behave interchangeably, whereas in our analysis they're different, and play distinct roles in the BPS combinations. The discrepancy seems to stem from the fact that \cite{Didenko2009} (as well as \cite{Bourdier2015,Iazeolla2011,Iazeolla2017}) dates from before the new homotopy prescription \cite{Gelfond2018,Didenko2018,Didenko2019,Gelfond2020} for the non-linear Vasiliev equations. It therefore cannot be trusted beyond linear order, where, without holography as a guide, type-A and type-B do \emph{seem} interchangeable (we thank Slava Didenko for a discussion on this point). From this point of view, the match between the type-A Didenko-Vasiliev solution and the boundary bilocal is a ``happy coincidence'', which doesn't necessarily generalize. Beyond this simplest case, we ought to trust the holographic dual.

The structure of the paper is as follows.
In section~\ref{sec:basics}, we briefly review our embedding-space formalism, HS algebra, the unfolded master field $C(x;Y)$ containing all HS field strengths, the Penrose transform between such master fields and twistor functions $F(Y)$, and the boundary theory.
In section~\ref{sec:typeBstory}, we construct the twistor description of boundary correlators for type-B, comparing to the type-A case. In section \ref{sec-3.SUSY}, we describe the $\mathcal{N}=2$ theory and its supersymmetry algebra. We then use this algebra to find the new $1/2$-BPS solutions. In Section \ref{sec:newbulkpicture} we discuss the new bulk interpretation mentioned above. Finally, section \ref{sec:discuss} is devoted to outlook.

\section{Basics and definitions}\label{sec:basics}

We consider parity-preserving type-B bosonic HS gravity on four-dimensional euclidean Anti-deSitter space. This theory contains exactly one field of every integer spin $s=0,1,2,3,...$ that interact via the Vasiliev equations~\cite{Vasiliev1990,VASILIEV1996,VASILIEV2000,Vasiliev2012}. In the following, we define our embedding space construction in which we will describe the linearized bulk theory, and give a short introduction into the higher-spin algebra. We then pack the linearized HS invariant field-strengths of all spins into a single master field and discuss how to generate solutions via a twistor transform. Finally, we describe the fermionic $U(N)$ vector model, which is the boundary dual to type-B theory under $AdS_4/CFT_3$. Notations used here are consistent with the ones used in~\cite{Neiman2017,David2020}, where the case of type-A was studied.

\subsection{Spacetime and boundary}\label{sec:spacetimeAndBoundary}

We describe the theory of type-B HS gravity on euclidean Anti-deSitter space ($EAdS_4$), for which we chose an embedding-space formalism. We define $EAdS_4$ as the future-pointing timelike hyperboloid in $\mathbb{R}^{1,4}$:
\begin{equation}
	EAdS_4=\left\{x\in\mathbb{R}^{1,4}\ |\ x_\mu x^\mu = -1,\; x^0>0\right\}\ ,\quad \text{Signature}(\eta_{\mu\nu})=(-,+,+,+,+)\ .
\end{equation}
We will use Greek letters $\mu, \nu, \ldots$ for indices in $\mathbb{R}^{1,4}$. Vectors in $EAdS_4$ are represented as elements of the tangent space of the hyperboloid at a point $x^\mu$, i.e. $v\cdot x\equiv v_\mu x^\mu =0$. Hence, we can form elements of the tangent space by acting with the projector $q_{\mu\nu}(x)=\eta_{\mu\nu}+x_\mu x_\nu$. Covariant derivatives in $EAdS_4$ are simply flat derivatives in $\mathbb{R}^{1,4}$ followed by the action of the projector on all indices:
\begin{equation}
	\nabla_\mu v_\nu = q_{\mu}^{\rho}(x)q_{\nu}^{\sigma}(x)\partial_\rho v_\sigma\ .
\end{equation}
Sometimes it is useful to have intrinsic coordinates. Choosing Poincar\'{e} coordinates for $EAdS_4$ we explicitly write
\begin{equation}\label{eq:intro:poincareCoords}
	dx^2=\frac{1}{z^2}(dz^2+d\bold{r}^2)\ ,\quad  x^\mu=\frac{1}{z}\left(\frac{1+z^2+r^2}{2},\bold{r},\frac{1-z^2-r^2}{2}\right)\ .	
\end{equation}
Since $EAdS_4$ is embedded as the hyperboloid, its asymptotic boundary is the projective lightcone in $\mathbb{R}^{1,4}$, i.e. a conformal 3-sphere $S_3$. A boundary point is represented as a lightlike vector $\ell^\mu$ with a rescaling equivalence relation $\ell^\mu \cong\lambda \ell^\mu$. We will use a concrete section of the $\mathbb{R}^{1,4}$ lightcone in order to obtain an intrinsic coordinate description. The section is defined by $\ell\cdot\ell_\infty = -\frac{1}{2}$, where $\ell_\infty$ is a fixed boundary point. In this section the boundary space is simply $\mathbb{R}^3$. This flat section follows from \eqref{eq:intro:poincareCoords} via the limit $z\rightarrow0$, $x^\mu\rightarrow\ell^\mu/z$,
\begin{equation}\label{eq:intro:flatSection}
	\ell^\mu=\left(\frac{1+r^2}{2},\bold{r},\frac{1-r^2}{2}\right)\ .
\end{equation}
We will use $i,j,k,...$ indices for intrinsic 3d boundary coordinates. The distance between two boundary points $\ell$ and $\ell'$ in this flat section is simply $|\bold{r}-\bold{r}'|=\sqrt{-2\ell\cdot\ell'}$. See \cite{Neiman2017} for further details on the relationship between intrinsic boundary and embedding-space quantities.\newline
One of the central elements that we will use here are the spinors of $\mathbb{R}^{1,4}$. Specifically, the spin-1/2 representation of $USp(2,2)$, which is the double cover of the $EAdS_4$ isometry group $SO(1,4)$. These 4-component Dirac-spinors are called twistors. We raise and lower their indices with the symplectic metric $I_{ab}$: 
\begin{equation}
	Y_a=I_{ab}Y^b\; ;\quad Y^a=Y_bI^{ba}\; ;\quad I_{ac}I^{bc}=\delta^b_a\ .
\end{equation}
A useful representation for $I_{ab}$ and the Dirac matrices $\left\{\gamma_\mu,\gamma_\nu\right\}=-2\eta_{\mu\nu}$ is given as
\begin{align}
	I_{ab}&=\left(\begin{array}{cc} 0 \ -i\sigma_2 \\ -i\sigma_2 \ 0 \end{array}\right)\ ,\\
	\left(\gamma^0\right)\indices{^a_b}=\left(\begin{array}{cc} 0 & 1 \\ 1 & 0 \end{array}\right)\ ,\quad
	\left(\gamma^4\right)\indices{^a_b}&=\left(\begin{array}{cc} 0 & -1 \\ 1 & 0 \end{array}\right)\ ,\quad
	\left(\gamma^k\right)\indices{^a_b}=\left(\begin{array}{cc} -i\sigma_k & 0 \\ 0 & i\sigma_k \end{array}\right)\ .\label{eq:intro:5dgamma_matrices}
\end{align}
Additionally, we will make use of the anti-symmetric product of two $\gamma$-matrices
\begin{equation}
	\gamma_{ab}^{\mu\nu}\equiv\gamma^{[\mu}_{ac}\gamma\indices{^{\nu]c}_b}\ .
\end{equation}
As we will be writing a lot of expressions including vectors contracted with $\gamma$-matrices, we will be adopting a ``slash-less'' Dirac-slash notation. Vectors missing the spacetime index $\mu$ are understood to be contracted with a $\gamma$-matrix (this applies to $\mathbb{R}^{1,4}$-vectors $\xi^\mu$, but also to bulk $x^\mu$, or boundary coordinates $\ell^\mu$). Furthermore, index contraction between spinors will be left implicit, with a down-to-up ordering; for example
\begin{equation}
	x\equiv\gamma\indices{_\mu^a_b}x^\mu\ ,\quad UAV\equiv U_a{A}\indices{^a_b}V^b\ .
\end{equation}
A dot will be used to indicate spacetime index contraction: $\ell\cdot \ell'=\ell_\mu\ell'^\mu$. Further relations in these notations and useful identities for the 5d $\gamma$-matrices can be found in \cite{Neiman2017,David2020}.\newline

\noindent Since we want to integrate certain functions of twistor variables, we define the following measure on twistor space:
\begin{equation}
	d^4Y\equiv\frac{\epsilon_{abcd}}{4!(2\pi)^2}dY^adY^bdY^cdY^d\ .
\end{equation}
The normalization is chosen such that it absorbs any factors of $2\pi$ that frequently appear in Gaussian and delta-function integrals. By fixing a space-time point, we can break the isometry group $SO(1,4)$ down to $SO(4)$. This results in twistor space decomposing into two spinor spaces at a bulk point $x^\mu$:
\begin{equation}\label{eq:intro:leftRightDecomposition}
	d^4Y=d^2y_Ld^2y_R\ ,\quad Y=y_L+y_R\ .
\end{equation}
In general, one can decompose a twistor into spinor components with respect to any pair of time- or lightlike $\mathbb{R}^{1,4}$-vectors $\xi\neq\xi'$, by defining the projectors
\begin{equation}\label{eq:intro:twistorDecomposition}
	P(\xi)=\frac{1}{2}\left(\sqrt{-\xi\cdot\xi}+\xi\right)\ ;\quad Y=\frac{2P(\xi)P(-\xi')Y}{\tr\left[P(\xi)P(-\xi')\right]}+\frac{2P(\xi')P(-\xi)Y}{\tr\left[P(\xi')P(-\xi)\right]}\ .
\end{equation}
Note that $P(\xi)$ and $P(-\xi)$ are orthogonal. We can define the above decomposition into left- and right-handed spinors at a bulk point $x^\mu$ by identifying $y_R=P(x)Y$, $y_L=P(-x)Y$. At a boundary point the two projectors $P(\ell)$ and $P(-\ell)$, while still orthogonal, are proportional to each other. Therefore, we cannot split twistor space into two orthogonal subspaces at a boundary point $\ell$. Instead, the measure splits as $d^4Y=-d^2y^*_\ell d^2y_\ell$, where the subspace $P(\ell)$ and quotient-space $P^*(\ell)$ are defined as
\begin{equation}
	y^a\in P(\ell)\Leftrightarrow ly=0\ ,\quad (y^*)^a\in P^*(\ell)\Leftrightarrow (y^*)^a\cong (y^*)^a+y^a \;\;\;\forall y^a\in P(\ell)\ .
\end{equation}
For more details on this decomposition and the $P^*(\ell)$ space see \cite{Neiman2017}.

\subsection{The higher-spin algebra}

The higher-spin algebra is an extension of the spacetime isometry group, which includes a set of additional generators corresponding to higher-spin fields. We represent this algebra in the star-product language~\cite{VASILIEV2000}. We define the star-product as a product acting on the twistor $Y_a$, as
\begin{equation}
	Y_a\star Y_b = Y_aY_b+2iI_{ab}\ .
\end{equation}
This product may be generalized to more general functions via an integral formulation~\cite{VASILIEV2000}:
\begin{equation}
	f(Y)\star g(Y)=\int d^4Ud^4Vf(Y+U)g(Y+V)e^{-iUV}\ .
\end{equation}
A useful identity is the star-product with a single twistor $Y_a$. In our notation it results in
\begin{align}
	\begin{split}\label{eq:intro:yStarRelation}
		Y_a\star f(Y)&=\left(Y_a-i\frac{\partial}{\partial Y^a}\right)f(Y)\ ,\\
		f(Y)\star Y_a&=\left(Y_a+i\frac{\partial}{\partial Y^a}\right)f(Y)\ .
	\end{split}
\end{align}
This product also admits a super-trace operation. The trace is cyclic for even functions and in general is given by
\begin{equation}
	\tr_\star\left[f(Y)\right]=f(0)\ ,\quad \tr_\star\left[f(Y)\star g(Y)\right]=(-1)^{\pi_f\pi_g}\tr_\star\left[g(Y)\star f(Y)\right]\ ,
\end{equation}
where the degree of parity $\pi_f$ is defined as $f(-Y)=(-1)^{\pi_f}f(Y)$. We will now use the above star-product to describe the free HS field equations.

\subsection{The Penrose-transform and linearized HS gravity}\label{sec:intro:penrosetransf}

The full Vasiliev equations are the topic of many reviews \cite{VASILIEV1996,VASILIEV2000,Vasiliev2012,Didenko2014,Bekaert2005,Bekaert2010}. Here, we will only require the linearized version, which encompasses the free field equations for all spin-$s$ fields. Note that it is possible to describe the linearized field content completely in terms of gauge-invariant field-strengths. If one were to consider interactions, the actual gauge-fields that define those field-strengths become important and one would need to introduce the additional fields which include the spin-$s$ gauge fields, their derivatives, and some auxiliary fields that make the explicit writing of the EOM possible. A full discussion can be found in the reviews \cite{Didenko2014,Bekaert2005}. Since we will describe bulk interactions only indirectly, via the boundary CFT, the linearized description given below will suffice.\newline

The field-strengths of the spin-$s$ fields include the spin-0 field $C^{(0,0)}(x)$; the spin-1 field-strength similar to the Maxwell-tensor in electrodynamics that splits into a left-handed and right-handed piece $C^{(2,0)}_{\alpha_1\alpha_2}$, $C^{(0,2)}_{\dot{\alpha}_1\dot{\alpha}_2}$; the Weyl-tensor for spin-2; and all the higher-spin generalizations. In general, we define the field-strengths split into left- and right-handed parts as
\begin{equation}
	\text{Spin-$s$ field-strengths: }\quad C^{(2s,0)}_{\alpha_1\ldots\alpha_{2s}}(x)\ ,\quad C^{(0,2s)}_{\dot{\alpha}_1\ldots\dot{\alpha}_{2s}}(x)\ .
\end{equation}
The intrinsic bulk spinor indices used here are just the SO(1,4) twistor indices contracted with projectors $P(\pm x)$ as defined in decomposition \eqref{eq:intro:leftRightDecomposition}. For example the right-handed spin-1 field-strength is written as
\begin{equation}
	C^{(0,2)}_{\dot{\alpha}\dot{\beta}}(x)=P(x)\indices{_{\dot{\alpha}}^a}P(x)\indices{_{\dot{\beta}}^b}C^{(0,2)}_{ab}(x)\ .
\end{equation}
In the linearized theory these fields satisfy free massless field equations on the $EAdS_4$ background (spin-0 is conformally coupled massless):
\begin{align}
	\text{spin = 0:}& & \left(\nabla_\mu\nabla^\mu+2\right)&C^{(0,0)}(x)=0 \ ; \label{eq:intro:freeEOM_spin0} \\
	\text{spin > 0:}& & \nabla\indices{^{\alpha_1}_{\dot{\beta}}} C^{(2s,0)}_{\alpha_1\ldots\alpha_{2s}}(x)=0 \ ; \quad&\quad\nabla\indices{_{\beta}^{\dot{\alpha}_1}} C^{(0,2s)}_{\dot{\alpha}_1\ldots\dot{\alpha}_{2s}}(x)=0\ . \label{eq:intro:freeEOM_spins}
\end{align}

In order to write the EOM in Vasiliev's unfolded formalism, we have to augment these field-strengths with an infinite set of auxiliary fields that represent their derivatives:
\begin{align}
	\begin{split}
		{C^{(2s+k,k)}}\indices{_{\alpha_1\ldots\alpha_{2s}\beta_1\ldots\beta_k}^{\dot{\beta}_1\ldots\dot{\beta}_k}}(x)=i^k\nabla\indices{_{(\beta_1}^{(\dot{\beta}_1}}\ldots\nabla\indices{_{\beta_k}^{\dot{\beta}_k)}}C\indices{_{\alpha_1\ldots\alpha_{2s})}}(x)\ ,\\
		{C^{(k,2s+k)}}\indices{^{\beta_1\ldots\beta_k}_{\dot{\beta}_1\ldots\dot{\beta}_k\dot{\alpha}_1\ldots\dot{\alpha}_{2s}}}(x)=i^k\nabla\indices{^{(\beta_1}_{(\dot{\beta}_1}}\ldots\nabla\indices{^{\beta_k)}_{\dot{\beta}_k}}C\indices{_{\dot{\alpha}_1\ldots\dot{\alpha}_{2s})}}(x)\ .
	\end{split}
\end{align}
The normalization factors of $i$ are introduced for convenience. We now have a single field corresponding to every integer spin representation of $SO(1,4)$, i.e. we have a field $C^{(m,n)}$ for every even $n+m$. We combine all these fields into a $0$-form master field. This is done by contracting the open indices with an auxiliary twistor, which is decomposed into the $SU(2)_L \otimes SU(2)_R$ spinor spaces at a bulk point $x$:
\begin{align}
	\begin{split}\label{eq:intro:masterfieldDef}
		C(x;Y)&=\sum_{m,n}\frac{1}{n!m!}C^{(m,n)}_{\alpha_1\ldots\alpha_m\dot{\alpha}_1\ldots\dot{\alpha}_n}(x)y_L^{\alpha_1}\ldots y_L^{\alpha_m}y_R^{\dot{\alpha}_1}\ldots y_R^{\dot{\alpha}_n}\ ,\\
		C^{(m,n)}_{\alpha_1\ldots\alpha_m\dot{\alpha}_1\ldots\dot{\alpha}_n}(x)&=P(-x)\indices{^{a_1}_{\alpha_1}}\ldots P(-x)\indices{^{a_m}_{\alpha_m}}P(x)\indices{^{a_{m+1}}_{\dot{\alpha}_{1}}}\ldots P(x)\indices{^{a_{m+n}}_{\dot{\alpha}_{n}}}\frac{\partial^{m+n}C(x;Y)}{\partial Y^{a_1}\ldots\partial Y^{a_{m+n}}}\bigg\vert_{Y=0}\ .
	\end{split}
\end{align}
Using the star-product, we can rewrite the free field equations in the simple form~\cite{Neiman2017}:
\begin{equation}\label{eq:intro:masterfieldEOM}
	\nabla_\mu C(x;Y)=C(x;Y)\star\left(\frac{i}{4}Y\gamma_\mu x Y\right)\ .
\end{equation}
This one equation fully encompasses the free field equations (\ref{eq:intro:freeEOM_spin0},~\ref{eq:intro:freeEOM_spins}). The fact that $C(x;Y)$ only contains bosonic fields is enforced by requiring it to be even in $Y$
\begin{equation}\label{eq:intro:masterfieldBosonicCondition}
	C(x;-Y)=C(x;Y)\ .
\end{equation}
At free field level, one can simply include half-integer spin fields by relaxing this condition~\cite{Didenko2014}.

At this point, we can introduce some special elements of the star-product algebra. \noindent One function that plays a special role in the following is the twistor delta-function $\delta(Y)$. It realizes a Fourier-transform under the star product
\begin{align}
	\delta(Y)\equiv\int d^4U e^{iUY},&\quad \int d^4Y f(Y)\delta(Y) = f(0)\ ,\\
	f(Y)\star\delta(Y)&=\int d^4Uf(U)e^{iUY}\ . \label{eq:f_star_delta}
\end{align}
As mentioned in the Introduction, $\delta(Y)$ will play the same algebraic role in our construction as the external Klein operator $k\bar k$ plays in \cite{Vasiliev1990,VASILIEV2000}. The key difference between the two is that multiplication by $k\bar k$ takes us outside the space of pure twistor functions $F(Y)$, whereas the product \eqref{eq:f_star_delta} remains within it (sending e.g. Gaussians to Gaussians).

In addition to delta-functions for the entire twistor space, we can also define functions that project onto the previously defined subspaces $P(\xi)$,
\begin{align}
	\delta_\xi(Y)&\equiv\int_{P(\xi)} d^2u\; e^{iuY}\ ,\label{eq:intro:generalDelta}\\
	\int_{P(\xi)} d^2u \,\delta_{\xi'}(u)f(u)&=\frac{2}{Tr[P(\xi)P(\xi')]}f(0)\ .
\end{align}
When $\xi^\mu$ is chosen as a bulk point $x^\mu\in EAdS_4$ we obtain the function $\delta_x(Y)$. These delta-functions play an important role in the higher-spin algebra. As is explained in detail in \cite{Neiman2017}, the adjoint action of $\delta_x(Y)$ implements the reflection $Y\rightarrow -xY$, whereas the adjoint action of $\delta(Y)$ realizes a $2\pi$ rotation. Concretely, when applied to a general twistor function $f(Y)$
\begin{align}
	\delta_x(Y)\star f(Y)\star \delta_x(Y)=f(-xY)\ ,\\
	\delta(Y)\star f(Y)\star \delta(Y)=f(-Y)\ ,\label{eq:intro:2pirot}
\end{align}
where we recall the condensed notation $(xY)^a\equiv x^\mu\gamma\indices{_\mu^a_b}Y^b$. Equation \eqref{eq:intro:2pirot} can be seen as a special case of the fact that two reflections along axes related by an angle $\alpha$ result in a rotation by $2\alpha$. In our particular case, $\delta_x(Y)\star\delta_{-x}(Y)=\delta(Y)$ simply becomes the statement that two reflections related by a $\pi$ rotation amount to a rotation by $2\pi$. Note also the relevance of equation \eqref{eq:intro:2pirot} to condition \eqref{eq:intro:masterfieldBosonicCondition} of $C(x;Y)$ being bosonic. We note further, that when acting in the HS fundamental, the map $\delta(Y)$ implements the antipodal map $x^\mu \rightarrow -x^\mu$. Indeed, when acting on the master field we obtain $C(x;Y)\star\delta(Y)=C(-x;Y)$, since $\delta_x(Y)\star\delta_x(Y)=1$ (see \eqref{eq:intro:penroseTransform} below). We will discuss the action of $\delta(Y)$ and $\delta_x(Y)$ when acting in the HS fundamental more in sections \ref{sec:susy_projectors} and \ref{sec:discuss}.

While delta-functions defined on subspaces of twistor space at a boundary point, like $\delta_\ell(Y)$, will be important below to describe higher-spin algebra elements corresponding to boundary single-trace primary operators, the function $\delta_{x}(Y)$ can be used to generate linearized bulk solutions: it implements the Penrose-transform. This Penrose-transform generates solutions to equation \eqref{eq:intro:masterfieldEOM}. Given a twistor function $F(Y)$ one can obtain a solution to the linearized bulk equations by taking a star-product with the delta-function \eqref{eq:intro:generalDelta} at a bulk point $x$
\begin{equation}\label{eq:intro:penroseTransform}
	C(x;Y)=F(Y)\star i\delta_x(Y)=i\int_{P(x)}d^2u\;F(Y+u)e^{iuY}\ .
\end{equation}
We note that this transformation is its own inverse up to sign. For details on how this version of the Penrose-transform relates to more commonly used notation see \cite{Neiman2017}.

\subsection{Boundary dual: The fermionic $U(N)$ vector model}\label{sec:intro:vectorModel}

As described in the Introduction, type-B theory is dual to the free fermionic $U(N)$ vector model. We will now describe the relevant features of this theory, and describe how it can be placed in our embedding space formalism of section \ref{sec:spacetimeAndBoundary}. The free fermionic $U(N)$ vector model consists of a spinor field $\psi^I$ with conformal weight $\Delta=1$. It lives on the conformal $3$-sphere that is the $EAdS_4$ asymptotic boundary, which is represented as the projective lightcone in $\mathbb{R}^{1,4}$. We will write most of the explicit formulae below in the flat section \eqref{eq:intro:flatSection}. The index $I=1,\ldots,N$ labels the internal $U(N)$ symmetry. We write the action as
\begin{equation}
	S=\int d^3r\; i\overline{\psi}_I\tau^k\partial_k\psi^I\ .
\end{equation}
Here, the 3d Dirac-matrices $\tau^k$ are simply Pauli-matrices up to a factor of $i$ to ensure the defining property
\begin{equation}
	\tau_i \equiv i\sigma_i\ ,\quad \left\{\tau_i,\tau_j\right\}=-2\delta_{ij}\ ,
\end{equation}
in alignment with the definitions of $\gamma_{\mu}$ in $\mathbb{R}^{1,4}$, as well as our choice of signature for $\eta_{\mu\nu}$. This theory contains an infinite tower of higher-spin conserved currents. By imposing conservation and tracelessness (or requiring them to be primary operators in the CFT) one obtains for $s\geq 1$~\cite{Anselmi1999}:
\begin{equation}\label{eq:typeB:higherSpinCurrents}
	J^{(s)}_{\psi k_1\ldots k_s}=\frac{1}{(2i)^s} \overline{\psi}_I\left(\sum^{s-1}_{m=0}(-1)^{m+1}\binom{2s}{2m+1}\tau_{(k_s}\overleftarrow{\partial}_{k_1}\ldots\overleftarrow{\partial}_{k_m}\overrightarrow{\partial}_{k_{m+1}}\ldots\overrightarrow{\partial}_{k_{s-1})}-\mbox{traces}\right)\psi_I\ ,
\end{equation}
where we have chosen a similar normalization as in the type-A version presented in~\cite{Neiman2017}, with factors of $i$ ensuring reality. The ``$-$ traces'' part indicates that all possible traces are to be subtracted, leaving us with completely symmetric traceless currents. We can hide away the open indices by contracting them with a polarization null-vector $\lambda^i$
\begin{equation}
	J_\psi^{(s)}(\bold{r},\boldsymbol{\lambda})\equiv J^{(s)}_{\psi k_1\ldots k_s}(\bold{r})\lambda^{k_1}\ldots\lambda^{k_s},\;\quad\mbox{with}\;\lambda_i\lambda^i=0\ .
\end{equation}
Note that, in general, mixed-symmetry currents which include more than one $\tau$-matrix could appear. However, in our case of three dimensions, where the $\tau$ are simply Pauli-matrices, those currents are related to the fully symmetric ones (see \cite{Grigoriev2018} for the case of general dimension, where this simplification does not occur). The currents $J_\psi^{(s)}(\bold{r},\boldsymbol{\lambda})$ have conformal weight $\Delta=s+1$, and with the addition of a spin-0 primary operator $J^{(0)}_\psi(\bold{r})=\overline{\psi}_I\psi^I$ they form the complete set of single-trace primary operators. The main difference to the type-A case of \cite{Neiman2017}, that appears at this level, is that the spin-0 operator $J_\psi^{(0)}(\bold{r})$ has conformal weight $\Delta=2$.

The propagator is defined as the usual inversion of the Dirac operator, which is obtained by simply applying the Dirac operator to the scalar propagator~\cite{Zhang2008}
\begin{align}
	i\slashed{\partial}S(\bold{r})=&-\delta(\bold{r})\ ,\\
	S(\bold{r})=&i\slashed{\partial}\left(-\frac{1}{4\pi|\bold{r}|}\right)=\frac{i}{4\pi}\frac{\slashed{r}}{|\bold{r}|^3}\ ,
\end{align}
where the slash indicates contraction with $\tau^k$. One obvious difference between this fermionic and the bosonic vector model as presented in \cite{Neiman2017}, is that this propagator carries spinor indices. This results in the expression for the correlators of currents to include a trace. For example the spin-1 $n$-point connected correlator takes the form
\begin{align}
	\begin{split}\label{eq:typeB:j1Correlator}
		\braket{J^{(1)}_\psi(\bold{r}_1,\boldsymbol{\lambda}_1)\ldots J^{(1)}_\psi(\bold{r}_n,\boldsymbol{\lambda}_n)}_c&= \\
		Ni^n(-1)\tr\Big(\prod_{p=1}^{n}&\slashed{\lambda}_pS(\bold{r}_p-\bold{r}_{p+1})+\text{perm}\Big)\ ,
	\end{split}
\end{align}
where ``$+$perm'' implies the addition of all cyclically in-equivalent permutations of boundary points $\bold{r}_p$.

\begin{figure}[h]
	\centering
	\begin{tabular}{c c c}
		\begin{tabular}{c}
			\begin{tikzpicture}
				\begin{feynman}
					\vertex [label=above:\(\bold{r}'_1\)](lp1) 	at (0,2);
					\vertex [label=above:\(\bold{r}_1\)](l1) 	at (2,2);
					\vertex [label=left:\(\bold{r}_4\)](l4) 	at (-1,1);
					\vertex [label=left:\(\bold{r}'_4\)](lp4) 	at (-1,-0.5);
					\vertex [label=right:\(\bold{r}'_2\)](lp2) 	at (3,1);
					\vertex [label=right:\(\bold{r}_2\)](l2) 	at (3,-0.5);
					\vertex [label=below:\(\bold{r}_3\)](l3) 	at (0,-1.5);
					\vertex [label=below:\(\bold{r}'_3\)](lp3) 	at (2,-1.5);
					
					\diagram* {
						(l4) -- [fermion, edge label=\(S\)] (lp1) -- [fermion, edge label=\(\slashed{\lambda}'_1S\slashed{\lambda}_1\)] (l1) -- [fermion, edge label=\(S\)] (lp2)  -- [fermion, edge label=\(\slashed{\lambda}'_2S\slashed{\lambda}_2\)] (l2),
						(l2) -- [fermion, edge label=\(S\)] (lp3) -- [fermion, edge label=\(\slashed{\lambda}'_3S\slashed{\lambda}_3\)] (l3) -- [fermion, edge label=\(S\)] (lp4) -- [fermion, edge label=\(\slashed{\lambda}'_4S\slashed{\lambda}_4\)] (l4),
					};
				\end{feynman}
			\end{tikzpicture}
		\end{tabular}&
		\begin{tabular}{c}$\longrightarrow$\end{tabular}&
		\begin{tabular}{c}
			\begin{tikzpicture}
				\begin{feynman}
					\vertex [label=above:\(\bold{r}'_1\)](lp1) 	at (0,2);
					\vertex [label=above:\(\bold{r}_1\)](l1) 	at (2,2);
					\vertex [label=left:\(\bold{r}_4\)](l4) 	at (-1,1);
					\vertex [label=left:\(\bold{r}'_4\)](lp4) 	at (-1,-0.5);
					\vertex [label=right:\(\bold{r}'_2\)](lp2) 	at (3,1);
					\vertex [label=right:\(\bold{r}_2\)](l2) 	at (3,-0.5);
					\vertex [label=below:\(\bold{r}_3\)](l3) 	at (0,-1.5);
					\vertex [label=below:\(\bold{r}'_3\)](lp3) 	at (2,-1.5);
					
					\diagram* {
						(l4) -- [fermion, edge label=\(S\)] (lp1) -- [scalar, edge label=\(\Pi\)] (l1) -- [fermion, edge label=\(S\)] (lp2)  -- [scalar, edge label=\(\Pi\)] (l2),
						(l2) -- [fermion, edge label=\(S\)] (lp3) -- [scalar, edge label=\(\Pi\)] (l3) -- [fermion, edge label=\(S\)] (lp4) -- [scalar, edge label=\(\Pi\)] (l4),
					};
				\end{feynman}
			\end{tikzpicture}
		\end{tabular}
	\end{tabular}
	\caption{This diagram shows position space connected correlators of spin-1 currents (left) and single-trace piece of the correlator of n-bilocals with sources $\Pi$ (right). When replacing every second propagator $S$ on the left, sandwiched with the polarization vectors $\slashed{\lambda}_p$, by a bilocal source, the correlator on the right is obtained. This action of ``dividing out'' propagators will be essential for obtaining the bilocal in section \ref{sec:typeBbilocal}.}
	\label{fig:spin1_bilocal_comparison}
\end{figure}
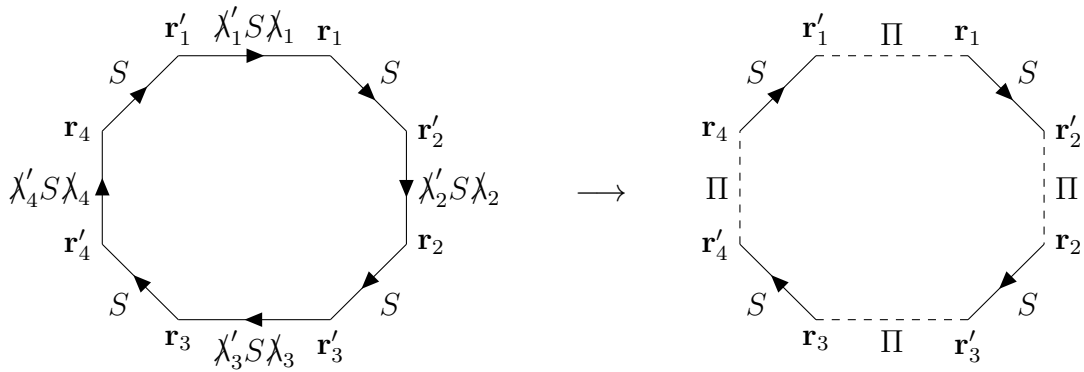

As was done in the type-A case, we want to repackage the infinite tower of currents~\eqref{eq:typeB:higherSpinCurrents} into a single composite operator, the bilocal
\begin{equation}\label{eq:intro:boundary_bilocal_def}
	{\mathcal{O}_\psi(\bold{r},\bold{r}')}\indices{^a_b}=\overline{\psi}_I^a(\bold{r})\psi^I_b(\bold{r}')\ .
\end{equation}
The bilocal includes enough information to encode all currents $J_\psi^{(s)}(\bold{r},\boldsymbol{\lambda})$. In fact, one can treat the bilocal as a fundamental field and discuss the free fermion in terms of it exclusively, as is done in \cite{Koch1996}. We, however, choose to keep it as a composite operator. In the AdS/CFT duality, bulk fields generate sources for the boundary currents. Instead of coupling the individual currents to sources, we use the bilocal description to couple to a single bilocal source $\Pi(\bold{r}',\bold{r})$
\begin{equation}\label{eq:typeB:action}
	S=\int d^3r\; i\psi_I\tau^k\partial_k\psi^I+\int d^3r\;d^3r'\;\tr\left[\mathcal{O}(\bold{r},\bold{r}')\Pi(\bold{r}',\bold{r})\right]\ .
\end{equation}
The partition function with action \eqref{eq:typeB:action} is Gaussian. Hence, we can rewrite it using the functional determinant over field space in the form (see \cite{Neiman2017, David2020} for analogous type-A)
\begin{equation}\label{eq:partitionfunction_bilocal}
	Z[\Pi]=\exp\left(-\frac{1}{N}\tr\left(\ln\left[1+\Pi G\right]-\Pi G\right)\right)\ .
\end{equation}
By expanding out the logarithm, we see that the partition function is build of single-trace pieces of the form $\tr\left[(\Pi G)^n\right]$. These single trace components can be evaluated, resulting in
\begin{equation}
	\begin{split}\label{eq:typeB:bilocalNPoint}
		\left<\tr\left[ O_\psi(\bold{r}_1,\bold{r}_1')\Pi(\bold{r}_1',\bold{r}_1)\right]\ldots\tr\left[ O_\psi(\bold{r}_n,\bold{r}_n')\Pi(\bold{r}'_n,\bold{r}_n)\right]\right>_c &= \\
		N(-1)\tr\Big(\prod_{p=1}^{n}\Pi(\bold{r}'_p,\bold{r}_p)S(\bold{r}_p-\bold{r}'_{p+1})&+\text{perm.}\Big)\ .
	\end{split}
\end{equation}
By comparing \eqref{eq:typeB:bilocalNPoint} and \eqref{eq:typeB:j1Correlator}, we see that the bilocal result is obtained from the connected correlator of spin-1 currents by replacing every second propagator $\slashed{\lambda}'_pS(\bold{r}'_p-\bold{r}_p)\slashed{\lambda}_p\rightarrow \Pi(\bold{r}'_p,\bold{r}_p)$ (please refer to figure~\ref{fig:spin1_bilocal_comparison}). We will exploit this similarity between spin-1 and bilocal correlators later, when we derive the twistor function corresponding to the bilocal $O_\psi(\bold{r},\bold{r}')$.\newline

We can rewrite the above in terms of our embedding-space language. Boundary vectors $v^i$ can be lifted to embedding-space $v^\mu$ via the pushforward defined by~\eqref{eq:intro:flatSection}. Conversely, an embedding-space vector $v^\mu$ at boundary point $\ell$ corresponds to a boundary vector if it is tangential to the lightcone $\ell\cdot v=0$ (subject to the equivalence $v^\mu\approx v^\mu + \alpha\ell^\mu$). These conditions are naturally enforced for boundary currents $J_\psi^{(s)}(\bold{r},\boldsymbol{\lambda})\rightarrow J_\psi^{(s)}(\ell(\bold{r}),\lambda)$ by replacing $\boldsymbol{\lambda}$ with its pushforward $\lambda^\mu=\lambda^i\partial_{r^i}\ell^\mu$ with $\lambda\cdot\lambda=0=\lambda\cdot\ell$ (see \cite{Neiman2017} for more details on lifting vectors and operators to embedding-space).

We can also describe the $3d$-conformal spinor field $\psi(\bold{r})$ in terms of an embedding-space spinor field $\psi(\ell)$ ($\Delta=1/2$) living in $\mathbb{R}^{1,4}$, which is an element of $P(\ell)$ i.e. $\ell\psi=0$ \cite{Iliesiu2016} (alternatively one could also choose it as an element of $P^*(\ell)$; our choice here will turn out to be more natural). Explicitly, in the chosen section \eqref{eq:intro:flatSection}, the 3d spinor is embedded as
\begin{equation}\label{eq:intro:spinorembedding}
	\psi_I^a(\ell)=\ell\left(\begin{array}{c}0\\\psi_I(\bold{r})^\alpha\end{array}\right)\ .
\end{equation}
We can find the embedding-space form of the fermion propagator by simply imposing $SO(1,4)$ invariance plus transversality conditions $\ell G_\psi(\ell,\ell')=0$, and $G_\psi(\ell,\ell')\ell'=0$. By choosing appropriate conformal weights and taking the normalization from before this results in
\begin{equation}\label{eq:typeB:embeddingPropagator}
	G_\psi(\ell,\ell')\indices{^a_b}=\frac{-i}{4\pi}\frac{(\ell\ell')\indices{^a_b}}{(-2\ell\cdot\ell')^{3/2}}\ .
\end{equation}
We will see this form of the propagator re-appear naturally from within our twistor space description later.

Similarly to the spinor field $\psi(r)$, we can lift the bilocal source into embedding-space coordinates. Since $\psi(r)$ was embedded in $P(\ell)$, we lift $\Pi(r',r)\indices{^i_j}$ to an element $\Pi(\ell',\ell)\indices{^a_b}$ with indices in $P^*(\ell')\otimes P^*(\ell)$.

\section{Twistor space and type-B, free fermion $AdS_4/CFT_3$}\label{sec:typeBstory}

In this section, we aim to describe the fermionic $U(N)$ vector model in  the same twistor space language that was used for the linearized bulk equations~\eqref{eq:intro:masterfieldEOM}. The case of type-A theory dual to the free boson was already considered in~\cite{Neiman2017}. Here, we apply similar methods to the analogous case of type-B theory. In particular, we will describe the correlators of boundary higher-spin currents in the bilocal language and identify the twistor function that corresponds to the bilocal operator $O_\psi(\ell,\ell')$. This bilocal can be used to extract all twistor functions corresponding to higher-spin currents as well as their correlators. This naturally includes the type-B spin-0 current whose correlators had to be separately calculated in the case of \cite{Didenko2013}.

We proceed as follows: after shortly reviewing the case of type-A \cite{David2020}, we derive the twistor functions that correspond to the boundary current operators \eqref{eq:typeB:higherSpinCurrents} in the higher-spin algebra, then we combine all currents into a single bilocal operator. This bilocal will naturally also include the operator corresponding to the scalar ``current'' of type-B theory, which a priori appears to be taking a special role due to its conformal weight $\Delta=2\neq s+1$. We will show that it naturally arises from the bilocal and can be easily included in the twistor space description.

\subsection{Twistor space description of the higher-spin current operators}

\subsubsection{Review of the type-A case}

In order to compute the type-B spin-$s$ boundary-to-bulk propagators and their Penrose-transform, it is useful to remind ourselves of the type-A case from \cite{Neiman2017,David2020}. Furthermore, these expressions will be useful in the next section \ref{sec-3.SUSY}. In \cite{David2020} the twistor functions corresponding to insertions of current operators $J_\phi^{(s)}(\ell,\lambda)$ for the bosonic $U(N)$ vector model dual to type-A theory were computed. These twistor functions take the form
\begin{equation}\label{eq:typeB:spinsTwistorFunctionTypeA}
	\kappa^{(s)}_\phi(\ell,\lambda;Y)\propto M^{a_1}\ldots M^{a_{2s}}\left(Y_{a_1}\ldots Y_{a_{2s}}+(-1)^s\frac{\partial^{2s}}{\partial Y^{a_1}\ldots\partial Y^{a_{2s}}}\right)\delta_\ell(Y)\ ,
\end{equation}
where the polarization spinors $M^a\in P^*(\ell)$ are related to the polarization vector $\lambda$ via
\begin{equation}
	(M\ell)^a(M\ell)^b=\gamma_{\mu\nu}^{ab}\ell^\mu\lambda^\nu\ .\label{eq:typeB:MpolDef}
\end{equation}
Note the special form of \eqref{eq:typeB:spinsTwistorFunctionTypeA}. The two terms in \eqref{eq:typeB:spinsTwistorFunctionTypeA} correspond to the left- and right-handed parts of the corresponding field-strength. We can see this explicitly by applying the Penrose-transform to each of the terms
\begin{align}
	(MY)^{2s}\delta_\ell(Y)\star\delta_x(Y) &\propto \frac{(M\ell y_L)^{2s}}{(\ell\cdot x)^{2s+1}}\exp\frac{iY\ell xY}{2\ell\cdot x}\ ,\label{eq:typeB:lefthandedKappaPhi}\\
	(M\partial_Y)^{2s}\delta_\ell(Y)\star\delta_x(Y) &\propto \frac{(M\ell y_R)^{2s}}{(\ell\cdot x)^{2s+1}}\exp\frac{iY\ell xY}{2\ell\cdot x}\ .\label{eq:typeB:righthandedKappaPhi}
\end{align}
We can extract  the HS field-strength from the RHS of \eqref{eq:typeB:lefthandedKappaPhi} via the definition of the master field \eqref{eq:intro:masterfieldDef}. We see that it contains a left-handed $C^{(2s,0)}(x)$ component, as well as all derivatives $C^{(2s+k,k)}(x)$ coming from the exponential. Similarly, \eqref{eq:typeB:righthandedKappaPhi} describes the right-handed spin-$s$ field-strength and its derivatives.

The key property of the twistor functions \eqref{eq:typeB:spinsTwistorFunctionTypeA} is that they compute the connected correlators of the boundary currents $J_\phi^{(s)}(\ell,\lambda)$, via the star-product expression
\begin{equation}
	\begin{split}\label{eq:typeB:typeAstarproductcorrelator}
		\braket{J^{(s_1)}_\phi(\ell_1,\lambda_1)\ldots J^{(s_n)}_\phi(\ell_n,\lambda_n)}_c&= \\
		\frac{N}{4}(-1)\tr_\star\Big(\kappa^{(s_1)}_\phi&(\ell_1,\lambda_1;Y)\star\ldots\star\kappa^{(s_n)}_\phi(\ell_n,\lambda_n;Y)+\text{perm.}\Big)\ ,
	\end{split}
\end{equation}
where ``$+$perm.'' refers to the addition of all cyclically inequivalent permutations of boundary-points $\ell_k$. Every star-product in \eqref{eq:typeB:typeAstarproductcorrelator} performs a Wick-contraction between the corresponding currents. In this way, \eqref{eq:typeB:typeAstarproductcorrelator} reproduces the CFT Feynman-diagrams element by element. As mentioned in the Introduction, eq. \eqref{eq:typeB:typeAstarproductcorrelator} is essentially a cleaner version of the star-product correlator formula from \cite{Didenko2012}, which used linearized bulk master fields $C(x;Y)$ rather than pure twistor functions $F(Y)$. In \cite{Didenko2012}, this was framed as a bulk expression for the boundary correlator; however, it is now widely understood that one must be very careful with this interpretation, because, as a bulk expression, it is extremely non-local. A bulk-local interpretation of \eqref{eq:typeB:typeAstarproductcorrelator} does exist \cite{Neiman2022}, in terms of (linearized) Didenko-Vasiliev ``black holes''. However, this interpretation won't be central to the present paper: we'll be content with considering bulk objects at the linearized level, as the Penrose transforms of the corresponding twistor functions. 

\subsubsection{Type-B construction}

Now let us turn to the case of type-B. We can guess the type-B equivalent of the type-A twistor functions $\kappa^{(s)}_\phi(\ell,\lambda;Y)$ by reminding ourselves of the following fact. The dictionary between the HS potentials (for $s>0$) and the components of the master field $C(x;Y)$ is different in type-A vs. type-B. The two dictionaries are related by a swapping of electric and magnetic parts of the field-strengths contained in $C(x;Y)$~\cite{Vasiliev2015}. From this we predict that the type-A twistor functions $\kappa^{(s)}_\phi(\ell,\lambda;Y)$, for $s>0$, are related to the type-B version $\kappa^{(s)}_\psi(\ell,\lambda;Y)$ by an electric/magnetic ``flipping''. In other words, the left- and right-handed parts of $\kappa^{(s)}_\phi(\ell,\lambda;Y)$ are rescaled by $\pm i$ in order to obtain the type-B version of these twistor functions. These additional factors of $\pm i$ were already observed by Giombi and Yin \cite{Giombi2011} on the level of boundary-to-bulk propagators. We therefore propose the type-B twistor functions that correspond to boundary current operators $J_\psi^{(s)}(\ell,\lambda)$ for $s>0$
\begin{equation}\label{eq:typeB:spinsTwistorfunction}
	\kappa^{(s)}_\psi(\ell,\lambda;Y)\propto M^{a_1}\ldots M^{a_{2s}}\left(Y_{a_1}\ldots Y_{a_{2s}}-(-1)^s\frac{\partial^{2s}}{\partial Y^{a_1}\ldots\partial Y^{a_{2s}}}\right)\delta_\ell(Y)\ ,
\end{equation}
with the normalization to be determined. Just as in the case of type-A, these twistor functions correspond to boundary operators $J_\psi^{(s)}(\ell,\lambda)$ on an element-to-element basis in Feynman-diagrams. As we will show, star-products of $\kappa^{(s)}_\psi(\ell,\lambda;Y)$ reproduce connected correlators of boundary currents $J_\psi^{(s)}(\ell,\lambda)$
\begin{equation}
	\begin{split}\label{eq:typeB:typeBstarproductcorrelators}
		\braket{J^{(s_1)}_\psi(\ell_1,\lambda_1)\ldots J^{(s_n)}_\psi(\ell_n,\lambda_n)}_c&= \\
		\frac{N}{4}(-1)\tr_\star\Big(\kappa^{(s_1)}_\psi&(\ell_1,\lambda_1;Y)\star\ldots\star\kappa^{(s_n)}_\psi(\ell_n,\lambda_n;Y)+\text{perm.}\Big)\ ,
	\end{split}
\end{equation}
where ``$+$perm.'' refers to the addition of all cyclically inequivalent permutations of boundary-points $\ell_k$. In particular, note that the exact same star-product expression reproduces the correlators in both cases \eqref{eq:typeB:typeAstarproductcorrelator},\eqref{eq:typeB:typeBstarproductcorrelators}, which in the case of type-B includes the minus sign from the fermion loop (as in \eqref{eq:typeB:j1Correlator}).

We will now explicitly perform the above calculation for the case of spin-$1$. This will allow us to fix the normalization of $\kappa_\psi^{(1)}(\ell,\lambda;Y)$, as well as show how the fermion-propagators emerge from the twistor-space expression. We will then use the spin-$1$ result, together with the observation from figure \ref{fig:spin1_bilocal_comparison}, to define the type-B bilocal. This is in complete analogy to type-A (see \eqref{eq:typeB:typeAbilocal}), where spin-0 was used instead of spin-1. As we will see, it is much easier to identify and divide out the propagator from the spin-1 expression in the type-B case. As is explained for the analogous case of type-A in \cite{David2020}, the twistor functions $\kappa_\psi^{(s)}(\ell,\lambda;Y)$ are contained as limits of derivatives of the bilocal.\newline

\noindent We will now show that the twistor function $\kappa_\phi^{(1)}(\ell,\lambda;Y)$ with left- and right-handed parts rescaled by $\mp i$ gives the correct type-B version of this function with our choice of normalization for $J_\psi^{(1)}(\ell,\lambda)$,
\begin{align}
	\kappa^{(1)}_\psi(\ell,\lambda;Y)=\frac{\pm1}{8\pi} M^{a_1}M^{a_2}&\left(Y_{a_1}Y_{a_2}+\frac{\partial^2}{\partial Y^{a_1}\partial Y^{a_2}}\right)\delta_\ell(Y)\ ;\label{eq:typeB:spin1_current_twistorf}\\[5pt]
	\begin{split}
		\braket{J^{(1)}_\psi(\ell_1,\lambda_1)\ldots J^{(1)}_\psi(\ell_n,\lambda_n)}_c&= \\
		\frac{N}{4}(-1)\tr_\star\Big(\kappa^{(1)}_\psi&(\ell_1,\lambda_1;Y)\star\ldots\star\kappa^{(1)}_\psi(\ell_n,\lambda_n;Y)+\text{perm.}\Big)\ .\label{eq:typeB:twistorSpin1CurrentCorrelator}
	\end{split}
\end{align}

First, we need to comment on the appearance of the sign-ambiguity in \eqref{eq:typeB:spin1_current_twistorf}. When taking star-products of three or more functions of the form \eqref{eq:typeB:spinsTwistorfunction}, we need to perform a complex Gaussian integration which has a sign-ambiguity depending on the analytical continuation from the real axis. This sign-ambiguity is essential for the higher-spin algebra and in general cannot be fixed. For a detailed discussion of this sign-ambiguity and how it relates to boundary locality, see \cite{David2020}. While on the level of correlators one can consistently fix this sign-ambiguity, we will keep the ambiguity apparent whenever it appears throughout this paper, similar to what was done originally in \cite{Neiman2017}.\newline

Let us start by computing the product of two $\kappa^{(1)}_\psi(\ell,\lambda;Y)$. In addition to integrals computed in \cite{Neiman2017,David2020}, we need integrals of the form ($\xi'\neq\pm\xi$)
\begin{equation}\label{eq:typeB:uvDeltaIntegral}
	\int_{P(\xi)}d^2u\int_{P(\xi')}d^2v\; u_{a_1}\ldots u_{a_n}v_{b_1}\ldots v_{b_k}e^{-iuv}\ .
\end{equation}
Without loss of generality we assume $k\geq n$. We will see that this integral vanishes except for the case $k=n$. We want to express this integral as derivatives on delta functions, and then perform a partial integration. A straightforward way to compute this, is by multiplying the integral by the expanded unity
\begin{equation}
	1=\frac{1}{2}\tr\left[P(\xi)P(\xi')\right]\int_{P(-\xi)}d^2w\;\delta_{-\xi'}(w)\ ,
\end{equation}
with $\delta_{-\xi'}(Y)$ as defined in \eqref{eq:intro:generalDelta}. We can then combine the measures\newline $d^2vd^2w=2/\tr\left[P(\xi)P(\xi')\right]d^4Z$ into a full twistor space measure and find
\begin{equation}
	\eqref{eq:typeB:uvDeltaIntegral}=i^n\int d^4Z\delta_{-\xi'}(Z)\left[\frac{2P(\xi')P(\xi)Z}{\tr\left[P(\xi)P(\xi')\right]}\right]_{b_1}\ldots\left[\frac{2P(\xi')P(\xi)Z}{\tr\left[P(\xi)P(\xi')\right]}\right]_{b_k}\frac{\partial^n}{\partial Z^{a_1}\ldots \partial Z^{a_n}}\delta_{\xi}(Z)\ .
\end{equation}
After partial integration, we can see that if any of the $2P(\xi')P(\xi)Z$ factors survive, they will vanish immediately after evaluating $\delta_{\xi}(Z)$, since this function projects the twistor $Z$ onto the orthogonal subspace $P(-\xi)$. Since $k\geq n$ the integral vanishes except for the case of $k=n$ in which case it evaluates to
\begin{equation}
	\eqref{eq:typeB:uvDeltaIntegral}=\delta_k^ni^n\left(\frac{2}{\tr\left[P(\xi)P(\xi')\right]}\right)^{n+1}\sum_{\sigma\in S_n}\prod_{i=1}^n\left[P(\xi)P(\xi')\right]_{a_{\sigma(i)}b_i}\ ,
\end{equation}
where the sum runs over all permutations $\sigma$ of indices $i\in\left\{1,\ldots,n\right\}$. This now allows us to compute the star-product of two $\kappa^{(1)}_\psi(\ell,\lambda;Y)$. We obtain
\begin{align}
	\begin{split}\label{eq:typeB:2PointKappa1}
		&\kappa^{(1)}_\psi(\ell_1,\lambda_1;Y)\star\kappa^{(1)}_\psi(\ell_2,\lambda_2;Y)=\\
		&\frac{M_1\ell_1\ell_2M_2}{(-2\ell_1\cdot\ell_2)^{\frac{3}{2}}}\frac{1}{4\pi^2(-2\ell_1\cdot\ell_2)^{\frac{3}{2}}}\left[M_2\ell_2\ell_1M_1+\frac{i}{\ell_1\cdot\ell_2}\left(M_1\ell_1\ell_2Y\right)\left(M_2\ell_2\ell_1Y\right)\right]\exp\frac{iY\ell_1\ell_2Y}{2\ell_1\cdot\ell_2}\ .
	\end{split}
\end{align}
In order to evaluate the star-product of three or more $\kappa^{(1)}_\psi(\ell,\lambda)$, we require the following twistor integral with symmetric matrix $A$
\begin{equation}
	\begin{split}
		\int_{P(\xi)}d^2u\; u_{a_1}\ldots u_{a_{2n}}e^{uAu/2}&=\\
		\frac{\pm1}{(\det_\xi (A))^{n+\frac{1}{2}}}&\frac{(2n)!}{2^nn!}\left(\left[P(\xi)AP(\xi)\right]_{(a_1a_2}\ldots\left[P(\xi)AP(\xi)\right]_{a_{2n-1}a_{2n})}\right)\ .
	\end{split}
\end{equation}
The determinant is the one defined in \cite{David2020}, where the simpler case of $n=0$ was evaluated. Note the emergence of a sign ambiguity, due to the complex Gaussian integral. With the help of this integral, we can find the $3$-point product
\begin{align}\label{eq:typeB:3PointKappa1}
	\begin{split}
		\kappa^{(1)}_\psi(\ell_1,&\lambda_1;Y)\star\kappa^{(1)}_\psi(\ell_1,\lambda_1;Y)\star\kappa^{(1)}_\psi(\ell_n,\lambda_n;Y)=\\
		&\frac{1}{(4\pi)^3}\left(\frac{M_1\ell_1\ell_{2}M_{2}}{(-2\ell_1\cdot\ell_{2})^{\frac{3}{2}}}\right)\left(\frac{M_2\ell_2\ell_3M_3}{(-2\ell_2\cdot\ell_3)^{\frac{3}{2}}}\right)\times\\
		&\frac{4}{(-2\ell_1\cdot\ell_3)^{\frac{3}{2}}}\left[M_3\ell_3\ell_1M_1+\frac{i}{\ell_1\cdot\ell_3}\left(M_1\ell_1\ell_3Y\right)\left(M_3\ell_3\ell_1Y\right)\right]\exp\frac{iY\ell_1\ell_3Y}{2\ell_1\cdot\ell_3}\ .
	\end{split}
\end{align}
Note that the final result has the same functional dependence in $Y$ as in the case for two $\kappa_\psi^{(1)}(\ell,\lambda;Y)$ \eqref{eq:typeB:2PointKappa1}. This is what results in the ``forgetfulness'' observed for boundary-to-bulk propagators in \cite{Didenko2013}. It is one of the important properties of the type-B bilocal, which will be constructed from $\kappa_\psi^{(1)}(\ell,\lambda;Y)$.\newline
\indent Due to the unchanged $Y$-dependence, we easily extrapolate to the $n$-point product 
\begin{align}\label{eq:typeB:nPointKappa1}
	\begin{split}
		\kappa^{(1)}_\psi(\ell_1,&\lambda_1;Y)\star\ldots\star\kappa^{(1)}_\psi(\ell_n,\lambda_n;Y)=\\
		&\frac{1}{(4\pi)^n}\left(\prod_{k=1}^{n-1}\frac{M_k\ell_k\ell_{k+1}M_{k+1}}{(-2\ell_k\cdot\ell_{k+1})^{\frac{3}{2}}}\right)\times\\
		&\frac{4}{(-2\ell_1\cdot\ell_n)^{\frac{3}{2}}}\left[M_n\ell_n\ell_1M_1+\frac{i}{\ell_1\cdot\ell_n}\left(M_1\ell_1\ell_nY\right)\left(M_n\ell_n\ell_1Y\right)\right]\exp\frac{iY\ell_1\ell_nY}{2\ell_1\cdot\ell_n}\ .
	\end{split}
\end{align}
Here we can identify factors of the embedding-space propagator \eqref{eq:typeB:embeddingPropagator} contracted with polarization spinors $M_k$. In fact, using the definition of the polarization spinor $M_k$ in terms of $\lambda_k$ \eqref{eq:typeB:MpolDef} we can compute directly (identifying index $n+1\hat{=}1$):
\begin{align}
	\begin{split}
		\prod_{k=1}^{n}\frac{(-i)(M_k\ell_k\ell_{k+1}M_{k+1})}{4\pi(-2\ell_k\cdot\ell_{k+1})^{\frac{3}{2}}}&=\\[5pt]
		=\left(\frac{-i}{4\pi}\right)^n&\frac{\tr\left[\ell_1\lambda_1\ldots\ell_n\lambda_n\right]}{[(-2\ell_1\cdot\ell_2)\cdots(-2\ell_n\cdot\ell_{1})]^{\frac{3}{2}}}\\
		=\tr\big[S(r_n&-r_1)\slashed{\lambda_1}S(r_1-r_2)\slashed{\lambda_2}\ldots S(r_{n-1}-r_n)\slashed{\lambda_n}\big]\ .
	\end{split}
\end{align}
Finally, taking the super-trace of \eqref{eq:typeB:nPointKappa1} gives the missing $n$-th fermion-propagator. This is how \eqref{eq:typeB:nPointKappa1} reproduces \eqref{eq:typeB:j1Correlator}.

\subsection{The type-B bilocal}\label{sec:typeBbilocal}

We now want to identify the twistor-space function that corresponds to the bilocal boundary operator $\mathcal{O}_\psi(\ell,\ell')$ from \eqref{eq:intro:boundary_bilocal_def}. As we saw in figure \ref{fig:spin1_bilocal_comparison}, correlation functions for the type-B bilocal can be obtained from correlation functions of spin-$1$ currents by removing certain propagators. This is completely analogous to the situation in type-A theory, as discussed in \cite{Neiman2017}, where the type-A bilocal twistor-function $K_\phi(\ell,\ell';Y)$ was found from spin-0 boundary-to-bulk propagators by dividing out the bosonic propagator
\begin{equation}
	\begin{split}\label{eq:typeB:typeAbilocal}
		K_\phi(\ell,\ell';Y)\equiv\frac{\kappa_\phi^{(0)}(\ell;Y)\star\kappa_\phi^{(0)}(\ell';Y)}{G_\phi(\ell,\ell')}=\\
		\frac{1}{\pi\sqrt{-2\ell\cdot\ell'}}\exp\frac{iY\ell\ell'Y}{2\ell\cdot\ell'}\ ,
	\end{split}
\end{equation}
where $G_\phi(\ell,\ell')=-1/(4\pi\sqrt{-2\ell\cdot\ell'})$ is the embedding space form of the propagator of the bosonic vector model.\newline

For our case of type-B, we note that the trace over fermionic propagators present in \eqref{eq:typeB:j1Correlator} turns into multiplicative factors in the twistor-space expression \eqref{eq:typeB:2PointKappa1}. This allows us to easily perform the removal of propagators highlighted in figure \ref{fig:spin1_bilocal_comparison} in order to construct the type-B bilocal twistor function from spin-$1$ correlators. We divide by the intermediate propagator and obtain:
\begin{align}\label{eq:typeB:bilocal}
	\begin{split}
		&M_1K_\psi(\ell_1,\ell_2;Y)M_2\equiv\frac{\kappa^{(1)}_\psi(\ell_1,\lambda_1;Y)\star\kappa^{(1)}_\psi(\ell_2,\lambda_2;Y)}{M_1G_\psi(\ell_1,\ell_2)M_2}=\\
		&\frac{-i}{\pi(-2\ell_1\cdot\ell_2)^{\frac{3}{2}}}\left[M_1\ell_1\ell_2M_2-\frac{i}{\ell_1\cdot\ell_2}\left(M_1\ell_1\ell_2Y\right)\left(M_2\ell_2\ell_1Y\right)\right]\exp\frac{iY\ell_1\ell_2Y}{2\ell_1\cdot\ell_2}\ .
	\end{split}
\end{align}
This bilocal fulfills similar properties as the bosonic one. Its star-trace is proportional to the propagator and it fulfills a ``forgetful'' property:
\begin{align}
	\tr_\star\left[K_\psi(\ell,\ell';Y)\right]\indices{^a_b}&=4G_\psi(\ell,\ell')\indices{^a_b}\ ,\label{eq:typeB:traceofbilocal}\\
	K_\psi(\ell_1,\ell'_1;Y)\indices{^a_b}\star K_\psi(\ell_2,\ell'_2;Y)\indices{^c_d}&=G_\psi(\ell_2,\ell'_1)\indices{^c_b}\,K_\psi(\ell_1,\ell'_2;Y)\indices{^a_d}\ .\label{eq:typeB:forgetful}
\end{align}
From these properties it is clear to see, how the star-product performs the Wick-contractions to draw correct Feynman-diagrams. The $n$-point correlator of bilocals follows directly from these properties. The normalization is also guaranteed since we fixed it for spin-$1$ in our earlier calculation. We obtain for the connected piece of the correlator of bilocal operators coupled to sources $\Pi(\ell,\ell')$
\begin{equation}
	\begin{split}
		\left<\tr\left[ O_\psi(\ell_1,\ell_1')\Pi(\ell_1',\ell_1)\right]\ldots\tr\left[ O_\psi(\ell_n,\ell_n')\Pi(\ell'_n,\ell_n)\right]\right>_c &= \\
		\frac{N}{4}(-1)\tr_\star\Big(\tr\left[K_\psi(\ell_1,\ell_1';Y)\Pi(\ell'_1,\ell_1)\right]\star\ldots\star&\tr\left[K_\psi(\ell_n,\ell'_n;Y)\Pi(\ell'_n,\ell_n)\right]+\text{perm.}\Big)\ .
	\end{split}
\end{equation}
We have therefore successfully found the type-B bilocal.

\subsection{Spin-0 from the bilocal}

As was stated in section \ref{sec:intro:vectorModel}, the bilocal is a composite operator that includes all necessary information to reconstruct the individual spin-$s$ current operators. Specifically, we can act on both ends of the bilocal with the same derivatives as in the explicit current expression \eqref{eq:typeB:higherSpinCurrents}, then taking the coincidence limit for $\ell'\rightarrow\ell$ and contracting open indices with polarization spinors $M^a$ will result in $\kappa_\psi^{(s)}(\ell,\lambda;Y)$. However, there are some subtleties here. The coincidence limit turns out to be singular. One therefore needs to take the Penrose-transform first, then apply the limit and transform back. The completely analogous case of type-A is explicitly calculated in \cite{David2020} for all spins. Here, we will do this calculation again, but for the special case of the type-B spin-0 field $\kappa_\psi^{(0)}(\ell;Y)$.

We start by computing the linearized bulk solution that is generated by the bilocal twistor-function \eqref{eq:typeB:bilocal}. It is given by the Penrose-transform which results in
\begin{equation}\label{bilocal_cpt}
	MK_\psi(\ell,\ell';Y)M'\star i\delta_{x}(Y)=(\pm i) MK_\psi(\ell,\tilde{\ell}';Y)\tilde{M}'\ ,
\end{equation}
where $\tilde{\ell}'\equiv-\ell'-2(x\cdot\ell)x$, and $\tilde{M}'\equiv ixM'$ are the reflections of boundary point $\ell'$ and spinor $M'$ along $x$. In \cite{Neiman2017} it is explained that the Penrose-transform can be seen as a ``square-root'' of a CPT reflection around a bulk point $x$, i.e. as the reflection applied to one ``leg'' of the bilocal. This interpretation matches precisely with the result \eqref{bilocal_cpt}. We will return to this fact in section \ref{sec:susy_projectors}.

We can now take the coincidence limit $\ell'\rightarrow\ell$, which results in the expression decomposing into two parts
\begin{align}
	\begin{split}\label{eq:typeB:bilocalLimit}
		\lim_{\ell'\rightarrow\ell}\big( M K_\psi(\ell&,\ell';Y)M'\star i\delta_{x}(Y) \big) =\\
		&(M\ell M')\frac{\mp i}{\pi(-2\ell\cdot x)^2}\left(1+\frac{iY\ell xY}{2\ell\cdot x}\right)e^{\textstyle\frac{iY\ell xY}{2\ell\cdot x}}+\\
		&\frac{\mp2}{(-2\ell\cdot x)^3}\left[(M\ell y_L)(M'\ell y_L)-(M\ell y_R)(M'\ell y_R)\right]e^{\textstyle\frac{iY\ell xY}{2\ell\cdot x}}\ .
	\end{split}
\end{align}
The second term, which contains the symmetric part, is exactly the Penrose-transform of the spin-1 operator $\kappa_\psi^{(1)}(\ell,\lambda;Y)$ when setting $M'=M$:
\begin{equation}
	\lim_{M'\rightarrow M}\lim_{\ell'\rightarrow\ell}\left(M K_\psi(\ell,\ell';Y)M'\star i\delta_{x}(Y) \right) = i\kappa_\psi^{(1)}(\ell,\lambda;Y)\star i\delta_x(Y)\ .
\end{equation}
This also functions as a check that our choices of normalization are consistent. It matches precisely the expectation from the boundary perspective, where the bilocal can be split into trace and symmetric part. Symmetrizing the coincidence limit of the embedding space bilocal gives us the spin-1 boundary current with the same proportionality factor of $i$ appearing:
\begin{equation}
	M\mathcal{O}_\psi(\ell,\ell)M=-\overline{\psi}_I(\bold{r})\slashed{\lambda}\psi^I(\bold{r})=iJ^{(1)}_\psi(\ell,\lambda)\ .
\end{equation}

We now turn to the first term in equation \eqref{eq:typeB:bilocalLimit}, i.e. the trace part of the bilocal in the coincidence limit. This can be identified as the boundary-to-bulk propagator of the type-B spin-0 field. We confirm directly via a boundary calculation that it corresponds to the spin-0 boundary operator $J_\psi^{(0)}(\ell)$
\begin{equation}
	\mathcal{O}_\psi(\ell,\ell)^{[ab]}=\frac{1}{2}\ell^{ab}\left(\overline{\psi}_I(\bold{r})\psi^I(\bold{r})\right)=\frac{1}{2}\ell^{ab}J_\psi^{(0)}(\ell)\ .
\end{equation}
Therefore, we are led to identify the Penrose-transform of the twistor function $\kappa_\psi^{(0)}(\ell;Y)$ from the coincidence limit of the bilocal twistor function
\begin{equation}
	\begin{split}
		\ell^{ab}\kappa_\psi^{(0)}(\ell;Y)\star i\delta_{x}(Y)&\equiv2\lim_{\ell'\rightarrow\ell}\left( K_\psi(\ell,\ell';Y)^{[ab]}\star i\delta_{x}(Y) \right) \\
		&=\ell^{ab}\frac{\mp 2i}{\pi(-2\ell\cdot x)^2}\left(1+\frac{iY\ell xY}{2\ell\cdot x}\right)e^{\textstyle\frac{iY\ell xY}{2\ell\cdot x}}\ .
	\end{split}
\end{equation} 
Inverting the Penrose-transform (by exploiting $\delta_x(Y)\star\delta_x(Y)=1$), gives us the result 
\begin{equation}\label{eq:typeB:spin0Twistor}
	\begin{split}
		\kappa_\psi^{(0)}(\ell;Y)&=\frac{\pm 1}{8\pi(\ell\cdot\ell_0)}Y\ell_0\frac{\partial}{\partial Y}\delta_{\ell}(Y)\\[5pt]
		&=\frac{\mp \ell_0^\mu}{4\pi(\ell\cdot\ell_0)}\lim_{\xi\rightarrow\ell}\sqrt{-\xi\cdot\xi}\,\partial_{\xi^\mu}\delta_{\xi}(Y)\\[5pt]
		&=\frac{\pm 1}{8\pi}\lim_{\xi\rightarrow\ell}\left(Y\partial_\xi\frac{\partial}{\partial Y}+\frac{2}{\sqrt{-\xi\cdot\xi}}\right)\delta_{\xi}(Y)\ ,
	\end{split}
\end{equation}
where we introduced an additional auxiliary boundary point $\ell_0$ not proportional to $\ell$. From the third expression it is clear that $\kappa_\psi^{(0)}(\ell;Y)$ is independent of $\ell_0$, and $\ell_0$ simply functions as a tool to write the expression in a simple form. Despite the limit, the second expression for $\kappa_\psi^{(0)}(\ell;Y)$ is often more useful than the first in direct computation of star-products. Due to its derivation from the bilocal, and the fixing of the normalization via the CFT perspective, it is guaranteed to reproduce connected correlators of spin-0 boundary operators
\begin{equation}
	\begin{split}
		\braket{J^{(0)}_\psi(\ell_1)\ldots J^{(0)}_\psi(\ell_n)}_c&= \\
		\frac{N}{4}(-1)\tr_\star\Big(\kappa^{(0)}_\psi&(\ell_1;Y)\star\ldots\star\kappa^{(0)}_\psi(\ell_n;Y)+\text{perm.}\Big)\ .\label{eq:typeB:twistorSpin0CurrentCorrelator}
	\end{split}
\end{equation}

We have therefore shown how to extract spin-0 from the bilocal. Furthermore, this shows that despite the special conformal weight $\Delta=2\neq s+1$ it is naturally contained in the bilocal and needs no special treatment. All other spins can be similarly extracted (as shown explicitly for type-A in \cite{David2020}) which also allows one to fix the normalization constants in \eqref{eq:typeB:spinsTwistorfunction} that we have not yet computed.\newline

Inspired by the above method for computing the bilocal by dividing out a propagator, we can find that if one contracts the bilocal $K_\psi(\ell,\ell';Y)\indices{^a_b}$ with the embedding-space description of the fermion propagator in $P^*(\ell)$: $-i\delta^b_a/4\pi \left(-2\ell\cdot\ell'\right)^{3/2}$, one exactly obtains the expression for the star-product of two $\kappa_\psi^{(0)}(\ell;Y)$:
\begin{equation}\label{eq:typeB:twoSpin0StarProduct}
	\begin{split}
		K_\psi(\ell,\ell';Y)\indices{^a_b}\frac{-i\delta_a^b}{4\pi \left(-2\ell\cdot\ell'\right)^{3/2}}
		&=\frac{-1}{2\pi^2(-2\ell\cdot\ell')^2}\left(1+\frac{iY\ell\ell'Y}{2\ell\cdot \ell'}\right)e^{\textstyle\frac{iY\ell\ell'Y}{2\ell\cdot\ell'}}\\
		&=\kappa_\psi^{(0)}(\ell;Y)\star\kappa_\psi^{(0)}(\ell';Y)\ .
	\end{split}
\end{equation}
Therefore, we want to identify the trace ($M_aM'^b\leftrightarrow \delta_a^b$) of the bilocal \eqref{eq:typeB:bilocal} as the ``spin-0'' bilocal (note that this still contains the full tower of integer spins)
\begin{equation}\label{eq:typeB:bilocalTrace}
	K_\psi^{tr}(\ell,\ell';Y)=\frac{1}{\pi\sqrt{-2\ell\cdot\ell'}}\left(1+\frac{iY\ell\ell'Y}{2\ell\cdot \ell'}\right)e^{\textstyle\frac{iY\ell\ell'Y}{2\ell\cdot\ell'}}\ ,
\end{equation}
where we have divided by a factor of $2i$ for later convenience. This expression will be important in the section \ref{sec-3.BPS}, where it forms an essential part of the new BPS black hole-like solutions.

\section{Internal supersymmetry in the higher-spin algebra} \label{sec-3.SUSY}

In this section we will demonstrate that our higher-spin covariant description of the $U(N)$ vector models dual to the type-A and type-B theory is embedded in an $\mathcal{N}=2$ supersymmetry algebra. Notably, this algebra is present in the already formulated star-product algebra, rather than something that relies on the addition of external fermionic oscillators or external Klein-operators as is usually done in supersymmetric HS theory \cite{VASILIEV2000,Sezgin2012} (see also \cite{Didenko2009,Sezgin2003}).

\subsection{Projectors for type-A and type-B}\label{sec:susy_projectors}

We start the discussion of the supersymmetry algebra by observing some interesting identities relating our type-A and type-B boundary descriptions. Namely, taking star-products of type-A and type-B twistor functions always results in the product vanishing
\begin{equation}
	\begin{split}
		K_\phi(\ell_1,\ell_2;Y)\star K_\psi(\ell_3,\ell_4;Y)&=0\\
		\Rightarrow \kappa^{(s)}_\phi(\ell,\lambda;Y)\star\kappa^{(s')}_\psi(\ell',\lambda';Y)&=0\ .
	\end{split}
\end{equation}
The fact that any star-product between them vanishes is thanks to the following projectors:
\begin{align}\label{eq:susy:typeABProjectors}
	P_\pm(Y)=\frac{1}{2}\left(1\pm\delta(Y)\right)\ .
\end{align}
They are projectors in the usual sense, i.e. $P_\pm(Y)\star P_\pm(Y)=P_\pm(Y)$, $P_\pm(Y)\star P_\mp(Y)=0$. These projectors were already written down in \cite{Neiman2017}, however their importance to the distinction of type-A and type-B was not yet realized there. We can get an intuitive understanding of the existence of these projectors by considering the role $\delta(Y)$ plays in the HS algebra. As we will see, $-\delta(Y)$, when acting in the HS fundamental, implements a rotation of {\it one} of the CFT-fields in the bilocal $\mathcal{O}_{\phi/\psi}(\ell,\ell')$ by $2\pi$. In the type-B case, this field is fermionic and the rotation produces a minus-sign. When adding this rotated version back to its original we obtain the projective property of $P_\pm(Y)$.

To understand why $-\delta(Y)$ acting in the HS fundamental implements a $2\pi$ rotation of one ``leg'', we remind the discussion in section \ref{sec:intro:penrosetransf}: when acting in the adjoint, $\delta(Y)$ implements a $2\pi$ rotation of the HS bulk field, which can be viewed as the combination of two reflections $\delta(Y)=\delta_x(Y)\star\delta_{-x}(Y)$. This decomposition of a $2\pi$ rotation into a reflection along $x$ and $-x$ can also be applied to the case of $\delta(Y)$ acting in the HS fundamental: recall from \eqref{bilocal_cpt} that, when acting in the fundamental (via multiplication on the right), $(\pm i)\delta_x(Y)$ reflects \emph{one} ``leg'' of the bilocal. Similarly, when acting from the left, it implements the reflection on the other ``leg'':
\begin{equation}
	i\delta_{x}(Y)\star MK_\psi(\ell,\ell';Y)M'=(\pm i) \tilde{M}K_\psi(\tilde{\ell},\ell';Y)M'\ .
\end{equation}
This is precisely the same behavior as was originally described for type-A in \cite{Neiman2017}. By repeated reflection around $x$ and $-x$ we get that $\delta_x(Y)\star\delta_{-x}(Y)=\delta(Y)$ acting in the fundamental rotates this same leg by $2\pi$, up to an undefined sign. As was discussed in \cite{Neiman2017}, it does not seem possible to consistently fix the sign ambiguity for the reflections $\pm i\delta_x(Y)$, and as a result the sign for the $2\pi$ rotation is left ambiguous. As shown below, by directly computing the action of the projectors $P_\pm(Y)$ on the bilocal, we see that $-\delta(Y)$ is the correct sign-choice for the $2\pi$ rotation.

Now, let us see how we can use these projectors to rediscover our type-B functions. First, let us notice the following two properties of the operator $\delta(Y)$:
\begin{align}
	\delta(Y)\star\delta_{\ell}(Y) &= \delta_{\ell}(Y)\star\delta(Y)=-\delta_{\ell}(Y) \ ; \\
	\delta(Y)\star Y_a &= -Y_a\star\delta(Y) \ ,
\end{align}
which lead to:
\begin{align}
	\label{eq:susy:pMinusIsTypeA}
	P_-(Y)\star\delta_{\ell}(Y)&=\delta_{\ell}(Y)\star P_-(Y)=\delta_{\ell}(Y)\ ,\\
	P_+(Y)\star\delta_{\ell}(Y)&=\delta_{\ell}(Y)\star P_+(Y)=0\ ;\\
	\label{eq:susy:starYChangesProjector}
	P_\pm(Y)\star Y_a &= Y_a\star P_\mp(Y)\ .
\end{align}
The first relation \eqref{eq:susy:pMinusIsTypeA} implies that we should assign the projection on $P_-(Y)$ to type-A. This is because $\delta_\ell(Y)\propto \kappa_\phi^{(0)}(\ell;Y)$ was identified in \cite{Neiman2017} as the twistor-function corresponding to the spin-0 boundary operator $J_\phi^{(0)}(\ell)$. As a result of the definition of the type-A bilocal \eqref{eq:typeB:typeAbilocal}, this implies that the entire type-A bilocal is in the $P_-(Y)$ space. Since all higher-spin current operators can be extracted from the bilocal (via derivatives and limits that commute with the star-product), they all lie in the space $P_-(Y)$. In fact, explicitly, we have:
\begin{align}
	\begin{split}
		\kappa^{(s)}_{\phi}(\ell,\lambda;Y) &\propto \frac{1}{2}\left[Y_{a_1}\ldots Y_{a_{2s}}+(-1)^s\frac{\partial^{2s}}{\partial Y^{a_1}\ldots\partial Y^{a_{2s}}}\right]\delta_{\ell}(Y) \\
		&= P_-(Y)\star\big(Y_{a_1}\ldots Y_{a_{2s}}\delta_{\ell}(Y)\big) = \big(Y_{a_1}\ldots Y_{a_{2s}}\delta_{\ell}(Y)\big)\star P_-(Y) \ .
	\end{split}
\end{align}

Secondly, we see from \eqref{eq:susy:starYChangesProjector} that taking star-products with $Y_a$ and any function that is in the $P_-(Y)$ space, such as $F_\phi(Y)=P_-(Y)\star F_\phi(Y)\star P_-(Y)$, flips the Projector on the side that $Y_a$ was applied to:
\begin{equation}
	F_\phi(Y)\star Y_a=F_\phi(Y)\star P_-(Y)\star Y_a=F_\phi(Y)\star Y_a\star P_+(Y)\ .
\end{equation}
Therefore, taking the star-product with $Y_a$ from the left and the right on a function in $P_-(Y)$ will generate a function that is fully in the $P_+(Y)$ projection. Doing this with the simple function $\delta_\ell(Y)$ and contracting the open indices with a spinor $M^a$ we obtain
\begin{equation}\label{eq:susy:kappa1Bder}
	M^aM^b(Y_a\star\delta_{\ell}(Y)\star Y_b)=M^aM^b\left(Y_aY_b+\frac{\partial^2}{\partial Y^a\partial Y^b}\right)\delta_{\ell}(Y)\ .
\end{equation}
This is simply the function we found for $\kappa^{(1)}_\psi(\ell,\lambda;Y)$! Resulting from this, the type-B bilocal \eqref{eq:typeB:bilocal} itself belongs to the $P_+(Y)$ space, and therefore all type-B higher-spin current operators as well. Explicitly, for $s>0$, we can write
\begin{align}
	\begin{split}
		\kappa^{(s)}_{\psi}(\ell,\lambda;Y) &\propto \frac{1}{2}\left[Y_{a_1}\ldots Y_{a_{2s}}-(-1)^s\frac{\partial^{2s}}{\partial Y^{a_1}\ldots\partial Y^{a_{2s}}}\right]\delta_{\ell}(Y) \\
		&= P_+(Y)\star\big(Y_{a_1}\ldots Y_{a_{2s}}\delta_{\ell}(Y)\big) = \big(Y_{a_1}\ldots Y_{a_{2s}}\delta_{\ell}(Y)\big)\star P_+(Y) \ ,
	\end{split}
\end{align}
while the spin-0 operator $\kappa_\psi^{(0)}(\ell;Y)$ can be extracted as the the anti-symmetric part of \eqref{eq:susy:kappa1Bder}:
\begin{equation}
	Y_{[a}\star\delta_{\ell}(Y)\star Y_{b]}\propto\ell_{ab}\kappa_\psi^{(0)}(\ell;Y)\ .
\end{equation}

It is important to mention one subtlety that we have so far ignored. Related to the sign-ambiguity that was mentioned in relation to expression \eqref{eq:typeB:spin1_current_twistorf}, the projectors $P_\pm(Y)$ can fail to be projectors. While this sign-ambiguity can be fixed consistently on the level of individual correlators \cite{David2020}, when considering more general boundary sources no consistent fixing of the ambiguity is possible. This breaking down of linearity also affects the projectors $P_\pm(Y)$ described above. For general linear combinations of fields they can fail to be well defined, which was shown explicitly in \cite{Neiman2017}. Since we only consider correlators and finite superpositions of fields here, this does not affect us, but in general needs to be kept in mind.

\subsection{The supersymmetry algebra}

Let us summarize some key observations made in the previous section:
\begin{itemize}
	\item we have two projectors that seem to separate the space of twistor functions into separate classes,
	\item taking products with $Y_a$ takes us from one to the other,
	\item and taking products with $Y_a$ increases the spin of the corresponding boundary operator (e.g. applying it once from each side took us from type-A spin-0 to type-B spin-1).
\end{itemize}
Based on these observations we can find the following algebra describing $\mathcal{N}=2$ supersymmetry. We define the charges
\begin{equation}\label{eq:susy:charges}
	Q^1_a=\frac{1+i}{2}Y_a,\quad Q^2_a=\frac{1-i}{2}\left(P_-(Y)-P_+(Y)\right)\star Y_a=-\frac{1+i}{2}\frac{\partial}{\partial Y^a}\delta(Y)\ .
\end{equation}
Normalizations are chosen such that their anti-commutator reproduces correctly the $SO(1,4)$ isometry generators of $EAdS_4$,
\begin{equation}
	\left\{Q_a^1,Q_b^1\right\}_\star=\left\{Q_a^2,Q_b^2\right\}_\star=iY_aY_b=-2\mathcal{M}_{ab}\ ,
\end{equation}
where the RHS is $(-2)$ times the twistor-index version of the $SO(1,4)$-generators $\mathcal{M}_{\mu\nu}=-\frac{i}{8}Y\gamma_{\mu\nu}Y$. This is what is expected from supersymmetry generators in constant curvature backgrounds~\cite{McKeon2003}. The generic anti-commutation relations also reveal the R-symmetry generator:
\begin{align}\label{eq:susy:susyalgebra}
	\left\{Q_a^r,Q_b^s\right\}_\star=iY_aY_b\delta^{rs}+2I_{ab}\epsilon^{rs}\big(\underbrace{\frac{i}{2}\delta(Y)}_{R(Y)}\big)\ ,\\
	\left[R(Y),Q_a^1\right]_\star=Q_a^2\ ,\quad\left[R(Y),Q_a^2\right]_\star=-Q_a^1
\end{align}

Now, let us see what kind of fermionic operators we can construct. So far we have spin-$s$ operators belonging to $P_-(Y)$ under multiplication from the left and right (i.e. $\kappa^{(s)}_\phi(\ell,\lambda;Y)$) and operators corresponding to $P_+(Y)$ on both sides ($\kappa^{(s)}_\psi(\ell,\lambda;Y)$). We can also construct objects which are ``type-A like'' on one side and ``type-B like'' on the other, by contracting with one $Q_a^r$. Starting with the simplest case of $\delta_{\ell}(Y)$ we obtain
\begin{align}\label{eq:susy:fermionTwistorFunction}
	&\delta_\ell(Y)\star Q_a^1=(1+i)P_-(Y)\star Y_a\delta_{\ell}(Y)\star P_+(Y)=\frac{1+i}{2}\left[Y_a+i\frac{\partial}{\partial Y^a}\right]\delta_{\ell}(Y)\ ,\\
	&Q_a^1\star\delta_\ell(Y)=(1+i)P_+(Y)\star Y_a\delta_{\ell}(Y)\star P_-(Y)=\frac{1+i}{2}\left[Y_a-i\frac{\partial}{\partial Y^a}\right]\delta_{\ell}(Y)\ .
\end{align}
Note that these twistor functions are odd in the twistor Y. They correspond to half-integer spin boundary operators. As mentioned earlier in this section, and in section \ref{sec:intro:penrosetransf}, the function $\delta(Y)$ implements $2\pi$-rotations when acting in the higher-spin adjoint: $\delta(Y)\star f(Y)\star\delta(Y)=f(-Y)$. For odd-functions of $Y$ this reproduces the expected minus sign, that fermions pick up under a $2\pi$-rotation.

Since we started with the boundary operator corresponding to $\overline{\phi}_I\phi^I$, and we saw that applying $Q_a^1$ on both sides takes us to $\kappa^{(1)}_\psi(\ell,\lambda;Y)\leftrightarrow i\overline{\psi}_I\slashed{\lambda}\psi^I$ we should think of the two fermionic elements as $\overline{\phi}_I\psi^J_a$ and $\overline{\psi}^b_J\phi^I$. In fact, these are precisely the operators they correspond to in the supersymmetric boundary theory. We can see that taking star-products of these fermionic elements produces exactly the expected elements: contracting the two operators does indeed reproduce the known bilocals, multiplied with a propagator that results from the contraction of the intermediate fields
\begin{align}
	M_1^aM_2^b\left[\delta_{\ell_1}(Y)\star Q_a^1\right]\star\left[ Q_b^1\star\delta_{\ell_2}(Y)\right]&=4\pi\frac{M_1\ell_1\ell_2M_2}{(-2\ell_1\cdot\ell_2)^{\frac{3}{2}}}K_\phi(\ell_1,\ell_2;Y)\ ,\\\
	M_1^aM_2^b\left[Q_a^1\star\delta_{\ell_1}(Y)\right]\star\left[\delta_{\ell_2}(Y)\star Q_b^1\right]&=\frac{4\pi i}{\sqrt{-2\ell_1\cdot\ell_2}}M_1K_\psi(\ell_1,\ell_2;Y)M_2\ .
\end{align}
The first line is a contraction of the twistor functions corresponding to the operators $(\overline{\phi}_I\psi^J_a)(\overline{\psi}^b_J\phi^I)$. The star-product implements a Wick-contraction of the $\psi$ fields, giving the fermion propagator and leaving the type-A bilocal remaining. In the second line the reverse order is computed leading to a contraction of the $\phi$ fields, resulting in a bosonic propagator multiplied with the type-B bilocal.

Because we can obtain both bilocals from the two fermionic elements \eqref{eq:susy:fermionTwistorFunction} they represent a minimal set. Using star-products of these elements, together with limits and derivatives of the bilocals, we can reproduce any correlation function of operators in the supersymmetric boundary CFT. This includes fermionic-bilocals that we can construct by contracting \eqref{eq:susy:fermionTwistorFunction} with $\delta_{\ell}(Y)$
\begin{align}\label{eq:susy:fermionicBilocals}
	(M_1Y)\delta_{\ell_1}(Y)\star\delta_{\ell_2}(Y)&=\frac{M_1\ell_1\ell_2Y}{(\ell_1\cdot\ell_2)^2}\exp\frac{iY\ell_1\ell_2Y}{2\ell_1\cdot\ell_2}\ ,\\
	\delta_{\ell_1}(Y)\star (M_2Y)\delta_{\ell_2}(Y)&=\frac{M_2\ell_2\ell_1Y}{(\ell_1\cdot\ell_2)^2}\exp\frac{iY\ell_1\ell_2Y}{2\ell_1\cdot\ell_2}\ .
\end{align}

\subsection{The BPS black hole-like solutions} \label{sec-3.BPS}

Having obtained a set of $\mathcal{N}=2$ supersymmetry charges, and being able to write solutions to the linearized bulk equations via the Penrose-transform, one can ask whether we can find a solution that preserves some of this symmetry. In the full supersymmetric theory based on external Klein-operators it has been shown that the Didenko-Vasiliev ``black hole'' is a BPS-solution \cite{Didenko2009}. We can reproduce this finding in our internal supersymmetry algebra, knowing that the linearized version of this BPS-solution is the type-A bilocal~\cite{David2020a}. Furthermore, we find an infinite tower of BPS-solutions that can be obtained from the type-A bilocal. They are a certain superposition of the type-A bilocal and higher orders of the helicity operator applied to it.

In the rest of this section, we will occasionally talk about ``type-A solutions'' or ``type-B solutions'' of the linearized bulk equations \eqref{eq:intro:masterfieldEOM}. The linearized equations themselves do not distinguish between type-A and type-B. Instead, ``type-A solutions'' and ``type-B solutions'' are shorthand for ``bulk duals to operators in the free-scalar boundary theory'' and ``bulk duals to operators in the free-fermion boundary theory''. As bulk master fields, the two types of solutions differ in their boundary behavior -- a subject we'll return to in detail in section \ref{sec:newbulkpicture}. 

\subsubsection{BPS combinations of boundary bilocals}

We proceed with proving that the type-A bilocal is $\frac{1}{2}$-BPS under the supersymmetry algebra of the previous subsection. We also show that a combination of type-A and traced type-B bilocal \eqref{eq:typeB:bilocalTrace} is just as BPS, but conserves the other ``half'' of supersymmetry generators. Finally, by realizing their relation to the helicity operator, we find an infinite tower of solutions.\newline

To find a BPS solution, we start by evaluating the adjoint action of generators \eqref{eq:susy:charges} on functions that are type-A or type-B. This can be done easily due to their dependence on the projectors $P_{\pm}(Y)$. Defining $F_\phi(Y)=P_-(Y)\star F_\phi(Y)\star P_-(Y)$, and $F_\psi(Y)=P_+(Y)\star F_\psi(Y)\star P_+(Y)$, using \eqref{eq:intro:yStarRelation} we obtain
\begin{align}\label{eq:susy:chargeAction}
	\begin{split}
		\left\{Q_a^1,F_\phi(Y)\right\}_\star&=(1+i)Y_a F_\phi(Y)\ ,\\
		\left\{Q_a^1,F_\psi(Y)\right\}_\star&=(1+i)Y_a F_\psi(Y)\ ,\\
		\left\{Q_a^2,F_\phi(Y)\right\}_\star&=(1+i)\frac{\partial}{\partial Y^a}F_\phi(Y)\ ,\\
		\left\{Q_a^2,F_\psi(Y)\right\}_\star&=-(1+i)\frac{\partial}{\partial Y^a}F_\psi(Y)\ .
	\end{split}
\end{align}
We want to find a combination $F(Y) = F_\phi(Y) + F_\psi(Y)$ that preserves as much of the $\mathcal{N}=2$ symmetry as possible, i.e. that is annihilated under some combination of SUSY-charges. To this end, we note the very similar form of the traced type-B bilocal \eqref{eq:typeB:bilocalTrace}, and the type-A bilocal \eqref{eq:typeB:typeAbilocal}. Both bilocals include exponentials of $YAY/2$ with $A\equiv i(\ell\ell'-\ell'\ell)/(2\ell\cdot\ell')$. In fact, even the quadratic prefactor of \eqref{eq:typeB:bilocalTrace} is of that form. Considering general combinations of the type-A and traced type-B bilocals results in the following two equations:
\begin{align}
	\begin{split}
		\left\{Q^1_a,K_\phi(\ell,\ell';Y)+bK_\psi^{tr}(\ell,\ell';Y)\right\}_\star&=\frac{(1+i)}{\pi\sqrt{-2\ell\cdot\ell'}}Y_a(1+b+bYAY/2)\exp(YAY/2)\ ,\\
		\left\{Q^2_a,K_\phi(\ell,\ell';Y)+bK_\psi^{tr}(\ell,\ell';Y)\right\}_\star&=\frac{(1+i)}{\pi\sqrt{-2\ell\cdot\ell'}}(AY)_a(2b-1+bYAY/2)\exp(YAY/2)\ .
	\end{split}
\end{align}
Since the matrix $A$ has full rank, no projectors exist to make any of the two lines vanish for themselves. Hence, the only option is to find a linear combination of $Q_1$ and $Q_2$ that vanishes. Because $Q^2_a$ acts via a derivative, we need to contract one of the charges with the matrix $A$, such that all four expressions in \eqref{eq:susy:chargeAction} become comparable (note $A^2=-1$). Furthermore, we can only find a vanishing superposition if the two polynomial factors in front of the exponential are equal. The only two solutions are $b=0$ and $b=2$:
\begin{align}
	&\begin{array}{l l}
		b=0:&K^{(0)}_{\text{BPS}}(\ell,\ell';Y)\equiv K_\phi(\ell,\ell';Y)\ ,\\
		&\left\{\left[Q^1-AQ^2\right]_a,K^{(0)}_{\text{BPS}}(\ell,\ell';Y)\right\}_\star=0\ ;
	\end{array}\\
	&\begin{array}{l l}
		b=2:&K^{(1)}_{\text{BPS}}(\ell,\ell';Y)\equiv K_\phi(\ell,\ell';Y)+2K^{tr}_\psi(\ell,\ell';Y)\ ,\\
		&\left\{\left[Q^1+AQ^2\right]_a,K^{(1)}_{\text{BPS}}(\ell,\ell';Y)\right\}_\star=0\ .
	\end{array}
\end{align}
We see that the type-A bilocal itself is $\frac{1}{2}$-BPS, confirming the finding of Didenko and Vasiliev \cite{Didenko2009} at linearized level in our language. The second new solution with $b=2$ is annihilated by the other ``half'' of supersymmetry generators.

At this point we make one further observation. As noted in the previous section, we can obtain type-B solutions from type-A by acting with the supersymmetry generators. When applying this to the type-A bilocal and tracing the result, we see that the type-B traced bilocal is actually obtained from the type-A one, by applying the helicity operator $H=\frac{1}{2}(Y^a\frac{\partial}{\partial Y^a}+2)$:
\begin{equation}\label{eq:susy:helicityIsQQ}
	K_\psi^{tr}(\ell,\ell';Y)=\frac{1}{2}Q^{1a}\star K_\phi(\ell,\ell';Y)\star Q^1_a=\frac{1}{2}\left(Y^a\frac{\partial}{\partial Y^a}+2\right)K_\phi(\ell,\ell';Y)\ .
\end{equation}
The fact that the action of the helicity operator switches between type-A and type-B is immediately clear, when we consider the discussion in the previous sections. The action of the helicity operator is twofold: 
\begin{itemize}
	\item when acting on a master field it rescales every spin by its value, hence changing the relative coefficients between the fields in the master field.
	\item it multiplies each helicity component by its sign. This switches the relative sign between left-handed and right-handed components, therefore switching between type-A and type-B.
\end{itemize}

So far we have found BPS solutions that are the $K^{(0)}_{\text{BPS}}(\ell,\ell';Y)=K_\phi(\ell,\ell';Y)$, and $K^{(1)}_{\text{BPS}}(\ell,\ell';Y)$, which is a superposition of $K_\phi(\ell,\ell';Y)$ and $HK_\phi(\ell,\ell';Y)$. We can find additional BPS-solutions by continuing this series and applying higher orders of the helicity operator. As we will show via a direct computation below, the full space of BPS-solutions is spanned by polynomials of $H+\frac{1}{2}$ acting on $K_{_\phi}(\ell,\ell';Y)$:
\begin{align}
	&K^{(\mathsf{N})}_{\text{BPS}}(\ell,\ell';Y)\equiv \left(H+\frac{1}{2}\right)^\mathsf{N}K_\phi(\ell,\ell';Y)\ , \label{eq:susy:BPS} \\
	&\left\{\left[Q^1-AQ^2\right]_a,K^{(\mathsf{N})}_{\text{BPS}}(\ell,\ell';Y)\right\}_\star=0\quad\text{when }\mathsf{N}\text{ is even}\ , \label{eq:susy:BPS_even} \\
	&\left\{\left[Q^1+AQ^2\right]_a,K^{(\mathsf{N})}_{\text{BPS}}(\ell,\ell';Y)\right\}_\star=0\quad\text{when }\mathsf{N}\text{ is odd}\ . \label{eq:susy:BPS_odd}
\end{align}
We can check the BPS conditions \eqref{eq:susy:BPS_even}-\eqref{eq:susy:BPS_odd} via a direct computation. From identities \eqref{eq:susy:chargeAction} we see that the action of $Q_a^1$ is simply the multiplication by a factor of $(1+i)Y_a$
\begin{equation}
	\left\{Q_a^1,K^{(\mathsf{N})}_{\text{BPS}}(\ell,\ell';Y)\right\}_\star=(1+i)Y_aK^{(\mathsf{N})}_{\text{BPS}}(\ell,\ell';Y)\ .
\end{equation}
On the other hand, $Q_a^2$ acts via a derivative, and results in a different sign depending on if the field is type-A or type-B. Expanding out $\left(H+\frac{1}{2}\right)^\mathsf{N}$ we can obtain
\begin{equation}
	\begin{split}
		\left\{Q_a^2,K^{(\mathsf{N})}_{\text{BPS}}(\ell,\ell';Y)\right\}_\star&=\left\{Q_a^2,\sum_{k=0}^\mathsf{N}\binom{\mathsf{N}}{k}\left(\frac{1}{2}\right)^{\mathsf{N}-k}H^kK_\phi(\ell,\ell';Y)\right\}_\star\\
		&=(1+i)\sum_{k=0}^\mathsf{N}\binom{\mathsf{N}}{k}\left(\frac{1}{2}\right)^{\mathsf{N}-k}(-1)^k\frac{\partial}{\partial Y^a}H^kK_\phi(\ell,\ell';Y)\\
		&=(1+i)\sum_{k=0}^\mathsf{N}\binom{\mathsf{N}}{k}\left(\frac{1}{2}\right)^{\mathsf{N}-k}(-1)^k\left(H+\frac{1}{2}\right)^k(-AY)_aK_\phi(\ell,\ell';Y)\\
		&=(-AY)_a(1+i)\sum_{k=0}^\mathsf{N}\binom{\mathsf{N}}{k}\left(\frac{1}{2}\right)^{\mathsf{N}-k}(-1)^k\left(H+1\right)^kK_\phi(\ell,\ell';Y)\\
		&=(-AY)_a(1+i)\left(-H-1+\frac{1}{2}\right)^\mathsf{N}K_\phi(\ell,\ell';Y)\\
		&=(-AY)_a(1+i)(-1)^\mathsf{N}K^{(\mathsf{N})}_{\text{BPS}}(\ell,\ell';Y)\ .\\
	\end{split}
\end{equation}
Remembering that $A^2=-1$ gives us immediately the result \eqref{eq:susy:BPS_even}-\eqref{eq:susy:BPS_odd}. We see that the distinction of $\mathsf{N}=$ even and $\mathsf{N}=$ odd comes from the way $Q^2_a$ acts on type-A versus type-B solutions. We therefore have of found two infinite sets of BPS-solutions that each respect half of the supersymmetry generators.

\subsubsection{Linearized bulk fields from the BPS boundary bilocals}

Let us now obtain the linearized bulk solutions that correspond to the BPS twistor functions \eqref{eq:susy:BPS}. As discussed above, these will be generalizations of the Didenko-Vasiliev solution, that combine both type-A and type-B fields. We thus find ourselves in a somewhat novel situation: both these sets of bulk fields are contained in the single twistor function \eqref{eq:susy:BPS}. The most straightforward way out is to first single out the type-A and type-B pieces of \eqref{eq:susy:BPS}, using e.g. the projectors $P_\pm(Y)$, and then apply the Penrose transform to these two pieces separately. However, since the function $\delta_x(Y)$ that realizes the Penrose transform \emph{commutes} with $P_{\pm}(Y)$, we can equivalently Penrose-transform into a single master field $C(x;Y)$ first, and separate it into type-A/type-B at the very end. This path will be more technically convenient, and we will follow it below. Its possible conceptual implications will be discussed in section \ref{sec:newbulkpicture}.

Let us then proceed to apply the Penrose transform \eqref{eq:intro:penroseTransform} to the BPS twistor functions \eqref{eq:susy:BPS}. The calculation for all values of $\mathsf{N}$ can be reduced to a single star-product, by realizing an alternative form of the helicity-operator $H$. The type-A bilocal is an exponential of the form $K^{(0)}_{BPS}(\ell,\ell';Y)\propto e^{YAY/2}$, with the matrix $A\equiv i(\ell\ell'-\ell'\ell)/(2\ell\cdot\ell')$. Repeatedly acting with the helicity-operator on such an exponential generates polynomial prefactors\footnote{The polynomial generated by the action of $H^n$ can be shown to be $P_n(YAY/2)=T_{n+1}(YAY/2)/(YAY/2)$, where $T_n(x)$ is the n-th Touchard polynomial~\cite{Touchard1939}. This follows from a simple induction proof using the recursion relation $x(1+\frac{d}{dx})T_{n-1}(x)=T_n(x)$, but will not be important here.}
\begin{equation}
	H^n e^{YAY/2}=P_n(YAY/2)e^{YAY/2}\ .
\end{equation}
By explicitly evaluating $H$ on functions of that form, we can quickly see that one can alternatively write $H^n$ using derivatives in an auxiliary variable. When acting on a function $f(YAY/2)$ that is an exponential with a polynomial prefactor
\begin{equation}
	H^nf(YAY/2)=\left(1+a\frac{d}{da}\right)^n f(aYAY/2)\big\vert_{a=1}\ ,
\end{equation}
where we substitute $a=1$ after taking all derivatives. Finally, we need to shift this operator by $1/2$ since BPS solutions are obtained by repeatedly acting with $H+\frac{1}{2}$, which simply adds a factor of $1/2$ to the auxiliary derivative term
\begin{equation}
	\left(H+\frac{1}{2}\right)^nf(YAY/2)=\left(\frac{3}{2}+a\frac{d}{da}\right)^n f(aYAY/2)\big\vert_{a=1}\ ,
\end{equation}
This lets us rewrite the Penrose-transform of the BPS-solutions as
\begin{equation}
	K^{(\mathsf{N})}_{\text{BPS}}(\ell,\ell';Y)\star i\delta_x(Y)=\frac{i}{\pi\sqrt{-2\ell\cdot\ell'}}\left(\frac{3}{2}+a\frac{d}{da}\right)^\mathsf{N}e^{aYAY/2}\int_{P(x)}d^2u\;e^{auAu/2+Du}\big\vert_{a=1}\ ,
\end{equation}
where $D=aYA-iY$. We can complete the square in order to obtain a Gaussian integral by choosing $u'=u+aP(x)AP(x)D/\det_x(A)$, where $\det_x(A)=1+2(\ell\cdot x)(\ell'\cdot x)/(\ell\cdot\ell')$. Then evaluating the integral (as solved in \cite{David2020}) leaves us with the bulk master field
\begin{equation}
	\begin{split}
		C_{\text{BPS}}^{(\mathsf{N})}(x&,\ell,\ell';Y)=K^{(\mathsf{N})}_{\text{BPS}}(\ell,\ell';Y)\star i\delta_x(Y)=\frac{\pm i}{\pi\sqrt{-2\ell\cdot\ell'-4(\ell\cdot x)(\ell'\cdot x)}}\times\\
		&\left(\frac{3}{2}+a\frac{d}{da}\right)^\mathsf{N}\frac{1}{a}\exp\left(\frac{(a^2+1)YAY}{4a\det_x(A)}-\frac{i}{4a(\ell\cdot\ell')det_x(A)}\right.\times\\
		&\big[(a^2-1)Y\ell\ell'xY-(a+1)^2(\ell'\cdot x)Y\ell xY+(a-1)^2(\ell\cdot x)Y\ell'xY\big]\bigg)\bigg\vert_{a=1}\ .
	\end{split}
\end{equation}
We can split into left- and right-handed components by setting $Y=y_L+y_R$. The part of the master field that contains all the left(right)-handed field strengths is then simply obtained by setting $y_R$($y_L$) to zero. This results in
\begin{align}
	\text{pref}\equiv&\frac{\pm i}{\pi\sqrt{-2\ell\cdot\ell'-4(\ell\cdot x)(\ell'\cdot x)}}\ ,\\
	\begin{split}
		C_{\text{BPS}}^{(\mathsf{N})}(x,\ell,\ell';y_L)&\equiv K^{(\mathsf{N})}_{\text{BPS}}(\ell,\ell';Y)\;\star\; i\delta_x(Y)\big\vert_{y_R=0}\\
		&=\text{pref}\left(\frac{3}{2}+a\frac{d}{da}\right)^\mathsf{N}\frac{1}{a}\exp\left(\frac{ay_LAy_L}{2\det_x(A)}\right)\bigg\vert_{a=1}\ ,
	\end{split}\\
	\begin{split}
		C_{\text{BPS}}^{(\mathsf{N})}(x,\ell,\ell';y_R)&\equiv K^{(\mathsf{N})}_{\text{BPS}}(\ell,\ell';Y)\star i\delta_x(Y)\big\vert_{y_L=0}\\
		&=\text{pref}\left(\frac{3}{2}+a\frac{d}{da}\right)^\mathsf{N}\frac{1}{a}\exp\left(\frac{y_RAy_R}{2a\det_x(A)}\right)\bigg\vert_{a=1}\ .
	\end{split}
\end{align}
Using definition \eqref{eq:intro:masterfieldDef} of the master field and expanding the exponential allows us to take the $a$-derivatives, and identify the spin-$s$ components as
\begin{align}\label{eq:susy:fullBPSfieldstrengths}
	C_{\text{BPS}\;\alpha_1\ldots\alpha_{2s}}^{(\mathsf{N})(2s,0)}(x,\ell,\ell')&=\text{pref}\;\frac{\left(\frac{1}{2}+s\right)^\mathsf{N}(2s)!}{s!}\left(\frac{[P(x)AP(x)]_{\alpha_i\alpha_{i+1}}}{2\det_x(A)}\right)^s\ ,\\
	C_{\text{BPS}\;\dot{\alpha}_1\ldots\dot{\alpha}_{2s}}^{(\mathsf{N})(0,2s)}(x,\ell,\ell')&=\text{pref}\;\frac{\left(\frac{1}{2}-s\right)^\mathsf{N}(2s)!}{s!}\left(\frac{[P(-x)AP(-x)]_{\dot{\alpha}_i\dot{\alpha}_{i+1}}}{2\det_x(A)}\right)^s\ ,
\end{align}
where indices are completely symmetrized. We obtain the expected result, namely the linearized Didenko-Vasiliev solution with extra powers of $(1/2\pm s)$ coming from $H+1/2$. We can now easily extract the type-A and type-B parts. We recall that $H$ takes us from the $P_\pm(Y)$ projection to $P_\mp(Y)$ (see \eqref{eq:susy:helicityIsQQ}). Therefore, we separate type-A from type-B by splitting the binomial $(H+\frac{1}{2})^N$ into even and odd terms in $H$, this leads in the equivalent $a$-operator to the expansion 
\begin{equation}
	\begin{split}
		\left(\frac{1}{2}+1+a\frac{d}{da}\right)^\mathsf{N}=&{}\frac{1}{2}\underbrace{\left[\left(\frac{3}{2}+a\frac{d}{da}\right)^\mathsf{N}+(-1)^\mathsf{N}\left(\frac{1}{2}+a\frac{d}{da}\right)^\mathsf{N}\right]}_{\text{even in }H}+\\
		&\frac{1}{2}\underbrace{\left[\left(\frac{3}{2}+a\frac{d}{da}\right)^\mathsf{N}-(-1)^\mathsf{N}\left(\frac{1}{2}+a\frac{d}{da}\right)^\mathsf{N}\right]}_{\text{odd in }H}\ .
	\end{split}
\end{equation}
As expected, for the type-A part (even in $H$) we obtain the ``electric'' part of \eqref{eq:susy:fullBPSfieldstrengths}
\begin{align}
	C_{\text{BPS},\phi\;\alpha_1\ldots\alpha_{2s}}^{(\mathsf{N})(2s,0)}(x,\ell,\ell')&=\text{pref}\;\frac{(2s)!\left[\left(\frac{1}{2}+s\right)^\mathsf{N}+\left(\frac{1}{2}-s\right)^\mathsf{N}\right]}{2s!}\left(\frac{[P(x)AP(x)]_{\alpha_i\alpha_{i+1}}}{2\det_x(A)}\right)^s, \label{eq:BPS_field_strengths_start} \\
	C_{\text{BPS},\phi\;\dot{\alpha}_1\ldots\dot{\alpha}_{2s}}^{(\mathsf{N})(0,2s)}(x,\ell,\ell')&=\text{pref}\;\frac{(2s)!\left[\left(\frac{1}{2}+s\right)^\mathsf{N}+\left(\frac{1}{2}-s\right)^\mathsf{N}\right]}{2s!}\left(\frac{[P(-x)AP(-x)]_{\dot{\alpha}_i\dot{\alpha}_{i+1}}}{2\det_x(A)}\right)^s.
\end{align}
Whereas type-B gives us the remaining ``magnetic'' part
\begin{align}
	C_{\text{BPS},\psi\;\alpha_1\ldots\alpha_{2s}}^{(\mathsf{N})(2s,0)}(x,\ell,\ell')&=\text{pref}\;\frac{(2s)!\left[\left(\frac{1}{2}+s\right)^\mathsf{N}-\left(\frac{1}{2}-s\right)^\mathsf{N}\right]}{2s!}\left(\frac{[P(x)AP(x)]_{\alpha_i\alpha_{i+1}}}{2\det_x(A)}\right)^s,\\
	C_{\text{BPS},\psi\;\dot{\alpha}_1\ldots\dot{\alpha}_{2s}}^{(\mathsf{N})(0,2s)}(x,\ell,\ell')&=\text{pref}\;\frac{-(2s)!\left[\left(\frac{1}{2}+s\right)^\mathsf{N}-\left(\frac{1}{2}-s\right)^\mathsf{N}\right]}{2s!}\left(\frac{[P(-x)AP(-x)]_{\dot{\alpha}_i\dot{\alpha}_{i+1}}}{2\det_x(A)}\right)^s. \label{eq:BPS_field_strengths_end}
\end{align}
We can extract the HS charges associated to these solutions, by directly comparing the prefactors in \eqref{eq:BPS_field_strengths_start}-\eqref{eq:BPS_field_strengths_end} with the derivation in \cite{David2020a}, where the charges of the linearized Didenko-Vasiliev solution itself were identified. We obtain:
\begin{align}
	Q^{(\mathsf{N},s)}_\phi&=\frac{i^s\sqrt{2^sN}}{4\pi\sqrt{-\ell\cdot\ell'}}\left[\left(\frac{1}{2}+s\right)^\mathsf{N}+\left(\frac{1}{2}-s\right)^\mathsf{N}\right]\ ,\\
	Q^{(\mathsf{N},s)}_\psi&=\frac{i^s\sqrt{2^sN}}{4\pi\sqrt{-\ell\cdot\ell'}}\left[\left(\frac{1}{2}+s\right)^\mathsf{N}-\left(\frac{1}{2}-s\right)^\mathsf{N}\right]\ , \label{eq:magnetic_charges}
\end{align}
(note that imaginary odd-spin charges are just the standard behavior in Euclidean signature -- c.f. the Euclidean Reissner-Nordstrom black hole).

\section{A new bulk picture}\label{sec:newbulkpicture}

As we have seen in sections \ref{sec:typeBstory}-\ref{sec-3.SUSY}, as far as boundary operators and correlators are concerned, the whole $\mathcal{N}=2$ supersymmetric theory (and in particular its type-A and type-B bosonic parts) is captured in a single space of twistor functions $F(Y)$. This raises a question: how should we think about the bulk fields and the Penrose transform \eqref{eq:intro:penroseTransform}? So far, in section \ref{sec-3.BPS}, we projected the twistor function into the type-A/type-B sectors, and interpreted the result as separate bulk fields, as usual. However, in the same calculation, we saw that it's actually more convenient to apply the Penrose transform \eqref{eq:intro:penroseTransform} to the entire twistor function first, and only separate the result into type-A/type-B at the very end. 

Perhaps a more natural interpretation is to drop the separation entirely. If we have a single space of twistor functions in the boundary picture, we can try to just interpret it directly as a single set of bulk fields, one for each spin, described by a single master field $C(x;Y)$. In this interpretation, type-A and type-B are no longer two separate sets of fields, but just different configurations of the \emph{same} set of fields. These configurations are distinguished by different boundary behavior of $C(x;Y)$: a different falloff rate for the $s=0$ component (conformal weight $\Delta=1$ vs. $\Delta=2$), as well as different leading components for $s>0$, related by an electric/magnetic duality. In the standard Vasiliev picture, this electric/magnetic ``flipping'' of the boundary conditions is canceled by a similar flipping in the dictionary between $C(x;Y)$ and the HS potentials. The standard picture thus contains two sets of bulk fields, both with magnetic boundary conditions, so that only the boundary conditions for $s=0$ differ. In contrast, if we take seriously the idea of type-A and type-B as different configurations of the same fields, then we should choose consistently one version of the $C(x;Y)$ $\leftrightarrow$ potentials dictionary -- for concreteness, the type-A one -- and accept the type-B boundary-to-bulk propagators as having non-standard boundary conditions (i.e. having fixed electric boundary data as in the second, spin-1, term in \eqref{eq:typeB:bilocalLimit}), to complement the standard ones of the type-A propagators. Of course, in the standard picture of HS holography, flipping boundary conditions leads to interacting boundary theories \cite{Klebanov2002,Sezgin2003,Giombi2013a,Giombi2013}. In our non-standard picture, the two boundary conditions instead correspond to two different \emph{free} boundary theories -- the bosonic one vs. the fermionic one -- whereas allowing \emph{both} boundary behaviors results in their supersymmetric union. All this is in conflict with the standard Vasiliev picture \cite{Vasiliev1990} (which has two separate sets of fields via the external Klein operators), as well as with standard AdS/CFT (which doesn't allow both kinds of boundary data of a bulk field to be dynamical at the same time). 

We do not yet know how to reconcile these tensions. However, they appear intriguingly related to a pair of separately important puzzles in holography: the doubled nature of the conformal boundary in de Sitter space, and the holographic role of the bulk self-dual sector. Let us start with the latter. Generally speaking, the self-dual sector of gauge and gravity theories is significantly simpler than the full theory. For HS gravity, the contrast is even sharper: the self-dual sector has an explicit bulk formulation in terms of physical degrees of freedom \cite{Skvortsov2018,Sharapov2022,Sharapov2022a}, while the full theory does not. But does the self-dual sector encode some important boundary observables? In asymptotically flat spacetime, the answer is yes: these are the MHV scattering amplitudes \cite{Bardeen1996,Rosly1997,Chalmers1996} (which, for self-dual HS theory, vanish \cite{Skvortsov2020}). However, in AdS, the answer is far less clear. Specifically, bulk self-duality (which sets the magnetic and electric boundary data equal to each other) is inconsistent with the standard boundary conditions of AdS/CFT (which freeze the magnetic data, while keeping the electric data dynamical). This has so far prevented self-dual bulk theories from playing a major role in holography (however, see \cite{Skvortsov2019}, where the self-dual theory was combined with its anti-self-dual mirror image to recover the boundary 3-point functions). With our new picture for HS holography, where the electric and magnetic data are both free to vary, the holographic role of the self-dual theory might be clarified: its holographic dual may turn out to be (an appropriate sector of) the \emph{supersymmetrized} boundary CFT.

We now turn to our story's possible relevance for de Sitter space. In AdS, freezing one of the two types of boundary data is necessary in order to construct normalizable states on the infinite spatial slices. But in (global) de Sitter, the spatial slices are finite, and the boundary is at infinite time, not at infinite distance; the two types of boundary data are just canonically conjugate operators on the asymptotic time slice, and so it makes little sense to freeze one of them. The question thus arises, what kind of ``doubled'' boundary CFT could describe this situation holographically. Our picture of the $\mathcal{N}=2$ HS theory may serve as an answer to this question. We stress that this is quite speculative: in the standard treatment, $\mathcal{N}=2$ supersymmetry in de Sitter leads to wrong kinetic-energy signs for the superpartners (in our case, the type-B fields); our hope is that this issue somehow won't arise in a picture where type-B is not a separate set of fields, but just a different boundary condition.

To further flesh out this relation, we note another intuition for why de Sitter allows both types of boundary data to be dynamical. This stems from the fact that de Sitter has \emph{two boundaries}: one in the past, and one in the future. In e.g. Euclidean AdS, where there is only one boundary, the two types of boundary data are tied to each other by a constraint: when evolving from the boundary inwards, the solution should ``meet itself'' correctly, i.e. without singularities, in the bulk. In de Sitter, this issue doesn't arise. When we evolve e.g. backwards in time from the future boundary, the solution is never forced to ``meet itself in the center'': instead of reaching a center, the spacetime re-opens into the past boundary. In the embedding-space language, this sense that de Sitter is ``twice larger'' than Euclidean AdS becomes very sharp: de Sitter is a one-sheeted hyperboloid, while Euclidean AdS is \emph{a single branch of} a two-sheeted hyperboloid. In particular, de Sitter is closed under the antipodal map $x\to -x$ (which interchanges its two boundaries), while Euclidean AdS gets sent into its other branch. This rhymes in several ways with the picture explored in this paper:
\begin{itemize}
	\item As mentioned in section \ref{sec:intro:penrosetransf}, the antipodal map is realized in HS algebra via multiplication by $\delta(Y)$ -- the same operation whose $\mp 1$ eigenvalues turn out to differentiate between type-A and type-B (see section \ref{sec:susy_projectors}).
	\item In the classification of bulk discrete symmetries, the antipodal map is of the CT type \cite{Neiman2014}. Up to an application of CPT, this is the same as parity P, which again differentiates between type-A (parity-even $C(x;Y)$) and type-B (parity-odd $C(x;Y)$). 
	\item For free massless fields (of all spins) in de Sitter space, the $\pm 1$ eigenvalues under the antipodal map precisely correspond to the two types of boundary data at infinity \cite{Neiman2014,Halpern2015}.
	\item In Vasiliev's approach to HS holography \cite{Vasiliev2012}, both types of boundary data are again dynamical (though in a different context), and this is again tied to a ``doubling'' of AdS via the antipodal map, expressed in \cite{Vasiliev2012} as $z\to -z$ in Poincare coordinates. 
\end{itemize}
These links deserve to be explored further. In the best-case scenario, HS holography and its supersymmetrization will lead us to a broadening of AdS/CFT that will be relevant for de Sitter, while also establishing contact with self-dual bulk computations.

\section{Outlook}\label{sec:discuss}

We have given a unifying twistor space description of boundary correlators and linearized bulk field solutions for type-B HS gravity. We identified the twistor function \eqref{eq:typeB:bilocal} corresponding to the boundary bilocal operator, which contains information about all HS current correlators. The spin-0 boundary ``current'' is naturally included in this bilocal description despite the special conformal weight $\Delta=2\neq s+1$. We can consistently reproduce all CFT current correlators from this bilocal description, which was done explicitly for spin-1, but holds in general in complete analogy to the type-A version of \cite{David2020}.

Furthermore, the combination of the type-A description of \cite{Neiman2017,David2020} and our type-B version here, has led to the discovery of projectors in the HS star-product algebra that can differentiate the two different sectors. We used those projectors to identify a global $\mathcal{N}=2$ supersymmetry algebra \eqref{eq:susy:susyalgebra}. An infinite set of $\frac{1}{2}$-BPS solutions were computed, which includes the type-A bilocal, identified previously as the linearized version of the Didenko-Vasiliev ``black hole'' \cite{Didenko2009,David2020}. The new solutions are generalizations of the Didenko-Vasiliev solution along the lines of \cite{Iazeolla2011,Iazeolla2017}, which can be generated by the action of the shifted helicity operator $H+\frac{1}{2}$, and carry different patterns of HS charges.

The new BPS solutions invite further study. As mentioned in the Introduction, the linearized Didenko-Vasiliev solution played an important role in a reformulation of the type-A bulk theory \cite{Lysov2022a,Neiman2022}. Perhaps the new BPS solutions can play a similar role (or even larger, as there are more of them) in the $\mathcal{N}=2$ theory. On a different note, the new solutions lend further weight to the open problem of revisiting the \emph{original} Didenko-Vasiliev solution in light of the amended prescription \cite{Gelfond2018,Didenko2018,Didenko2019,Gelfond2020} for the non-linear Vasiliev equations. If our new holographically-motivated solutions can be recovered in such an approach, this would constitute a strong consistency check between the Vasiliev equations and the holographic picture.

The main conceptual novelty of our formalism was the description of the whole $\mathcal{N}=2$ supersymmetric theory within a single space of twistor functions $F(Y)$. As discussed in section \ref{sec:newbulkpicture}, this suggests an intriguing (but confusing!) alternative bulk picture, where the type-A and type-B bosonic sectors are described by the same set of fields. It remains to be seen whether this picture is merely a curiosity, or whether it points to genuine insights about e.g. dS/CFT and holography for the self-dual sector.

\section{Acknowledgments}
We thank Per Sundell, Vyacheslav E. Didenko and Mirian Tsulaia for discussions. This work is supported by the Quantum Gravity Unit of Okinawa Institute of Science and Technology Graduate University (OIST).

%----------------------------------------------------------------------------------------
%	BIBLIOGRAPHY
%----------------------------------------------------------------------------------------

\providecommand{\href}[2]{#2}\begingroup\raggedright\endgroup

\end{document}